\documentclass[prb,twocolumn,superscriptaddress,showpacs]{revtex4}
\usepackage{graphicx}
\usepackage{float}
\usepackage{dcolumn}
\usepackage{bm}
\usepackage{mathrsfs}
\usepackage{ulem}
\usepackage{hyperref}
\usepackage{xcolor}
\usepackage{soul}
\usepackage{amsmath}
\usepackage{amssymb}
\usepackage{mathrsfs}
\usepackage{siunitx}

\begin{document}
\title{Phase solitons in a weakly coupled three-component superconductor }
\author{Yuriy Yerin}
\affiliation{ Dipartimento di Fisica e Geologia, Universitá degli Studi di Perugia, Via Pascoli, 06123 Perugia, Italy}
\author{Stefan-Ludwig Drechsler}
\affiliation{Institute for Theoretical Solid State Physics, Leibniz-Institut für Festkörper- und Werkstoffforschung IFW-Dresden, D-01169 Dresden, Helmholtzstraße 20}
\date{\today }

\begin{abstract}
Based on the phenomenological Ginzburg-Landau approach, we investigate phase solitonic states in a class of three-component superconductors for mesoscopic doubly-connected geometry (thin-walled cylinder) in external magnetic fields. Analysis of the Gibbs free energy of the system shows that solitonic states in a three-component superconductor are thermodynamically metastable and separated from the ground state by a sizable energy. Our results demonstrate that despite the presence of a zoo of phase solitons states the earlier proposed method [Phys. Rev. B 96, 144513 (2017)] for the detection of BTRS (broken time-reversal symmetry) in multiband superconductors remains valid and useful.
\end{abstract}

\maketitle

\section{Introduction}

The discovery of unconventional superconductivity with a multicomponent complex order parameter caused an exploding growth of activities in condensed matter. In particular, thereby the formation of a plethora of topological defects: phase kinks, phase domains, vortices that carry fractional magnetic flux values, and phenomena like the fractional Josephson effect might be allowed there (see Refs. \onlinecite{Mermin, Teo, Lin1, Tanaka1, Yerin1} and references therein). The complexity of the order parameter with the presence of several components may break the time-reversal symmetry \cite{Tafti, Watashige, Grinenko1, Grinenko2, Ghosh}. Such superconductors are expected to have a nontrivial response in external magnetic fields and special magnetic properties even at zero field,  i.e. spontaneous magnetic ordering. \cite{Gillis1, Huang1, Yerin2, Chubukov1, Yanagisawa1, Babaev1, Babaev2, Babaev3, Babaev4, Babaev5, Babaev6, Babaev7}. 

Ignoring a long history with various materials included, often the phenomenon of two-band and two-gap superconductivity has been considered in the context of a magnesium diboride ($\text{MgB}_{\text{2}}$) and the family of iron-based superconductors, where the presence of two gaps is a well established experimental fact. In this regard, it may be thought that the three-component superconductivity is nothing more than an academic model. However, experimental evidence of three-gap superconductivity was unambiguously found in some iron-pnictide compounds of 122 and 111 types \cite{Ding, Morozov, remark}. Moreover, enormous advances in a density-functional theory predict the possibility of the occurrence of three superconducting gaps in the strain-enhanced atomically thin $\text{MgB}_{\text{2}}$ and $\text{AlB}_{\text{2}}$-based films \cite{Bekaert, Zhao}. Furthermore, such a model is relevant too for a system of three Josephson junctions between three usual single-band bulk superconductors that form a prism \cite{Malomed}.

In our previous work Ref. \onlinecite{Yerin2} within the Ginzburg-Landau  formalism we have analyzed the homogeneous ground state of a three-component superconductor in a convenient for theoretical analysis geometrical form of a ring or tube in the presence of an external magnetic field. We have shown that depending on the inter-component coupling constants, a magnetic flux can induce current density jumps in such superconducting geometries that are related to transitions from broken time reversal symmetry (BTRS) to time-reversal symmetric (TRS) states and vice versa. Among the low energy excitations there will be also topological phase solitons, to be studied here \cite{Tanaka2002, Gurevich2003, Garaud, Lin2, Yerin3, Vakaryuk, Arisawa, Samokhin1, Vodolazov}. Such excitations are forbidden in the bulk due to divergent total energy in the spatially unlimited case. But they can have finite energy in special geometries limited in the transverse to the tube direction as realized for long tubes.

Moreover, the phase solitons in this geometry can be induced by an external parallel magnetic field \cite{Yerin3}. It should be noted that the observation of such topological defects arising from the interaction between two order parameters in quasi-one-dimensional superconducting rings consisting of two parallel layers with weak Josephson coupling has been already reported recently \cite{Bluhm} .We further note that the existence of phase solitons was verified experimentally via the observation of fractional vortices generated in a thin superconducting \textit{heterostructure} in the form of a Nb/Al-AlO${}_{x}$/Nb trilayer \cite{Tanaka2}. 

The detailed knowledge about these topological defects can be very helpful for the detection and assignment of the BTRS phenomenon. Thus, with this in mind, the goal of the present paper is to provide a comprehensive picture of possible inhomogeneous states like phase solitons for three-component superconducting systems with a doubly-connected geometry. Our paper is organized as follows.  In Sec. II, we present the model and the main equations of the Ginzburg-Landau approach for the description of phase solitons within a three-component system. In Sec. III, we provide analytical results for a qualitative insight in the structure of these topological defects, discuss their stability and find their energy.  The results of numerical investigations of phase solitons are shown in Sec. IV.  The general discussion concerning the properties and the possible experimental detection of phase solitonic states is found in Sec. V.  We summarize our conclusions in Sec. VI. Two appendices with the derivation of the governing equations and the particular examples of phase solitons are reported at the end of our paper.

\section{Model and basic equations}

\begin{figure}
\includegraphics[width=0.99\columnwidth]{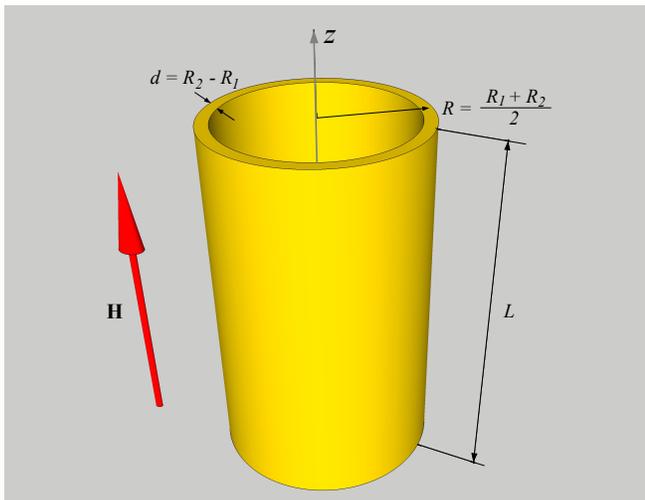}
\caption {The geometry of the problem (schematically) taken from Ref. \onlinecite{Yerin2}.}
\label{cylinder}
\end{figure}

As mentioned above we consider a Ginzburg-Landau functional for the free (Gibbs) energy, for a superconductor with a three-component superconducting order parameter in the form of a thin long tube with a thin wall (the thickness is supposed to be much smaller than the characteristic coherence length(es), while the radius has to be much larger), whose symmetry axis is the \textit{z} axis of cylindrical coordinates  ($r,\varphi ,z$). The constant external magnetic field $H$ is applied along the symmetry axis: $H=\left(0,0,H\right)$ (see Fig. (\ref{cylinder})). In such a situation we neglect the\textit{ r}-and \textit{z }dependencies of the superconducting order parameter, which are relevant for thick short tubes. These conditions preclude the formation of any vortices in the wall of the cylinder and guarantee that the self-induced magnetic fields are small. Noteworthy, similar results with some correction due to the finite demagnetization factor are also expected for thin rings, which will describe approximately experiments like in Refs. \onlinecite{Bluhm, Tanaka2}.

Thus, we start from the Gibbs free-energy functional of a superconducting cylinder. In view of the quasi-homogeneity along the \textit{z} axis, it takes the following approximate form:
\begin{equation}
\label{Gibbs_energy}
\mathbb{G} = {\mathbb{F}_1} + {\mathbb{F}_2} + {\mathbb{F}_3} + {\mathbb{F}_{\operatorname{int} }} + \int\limits_{{\Omega _c} + {\Omega _o}} {\frac{{{{\left( {{\text{rot }}{\mathbf{A}} - {\mathbf{H}}} \right)}^2}}}{{8\pi }}} d{\text{V}},
\end{equation}
where the partial contribution from the first component ${\mathbb{F}_1}$ has the form
\begin{equation}
\label{Gibbs_energy1}
{\mathbb{F}_1} = \int\limits_{{\Omega _c}} {\left[ {{\alpha _1}{{\left| {{\psi _1}} \right|}^2} + \frac{1}{2}{\beta _1}{{\left| {{\psi _1}} \right|}^4} + {{\bar \kappa }_1}{{\left| {{\mathbf{\Pi }}{\psi _1}} \right|}^2}} \right]} d{\text{V}},
\end{equation}
from the second  one ${\mathbb{F}_2}$ is
\begin{equation}
\label{Gibbs_energy2}
{\mathbb{F}_2} = \int\limits_{{\Omega _c}} {\left[ {{\alpha _2}{{\left| {{\psi _2}} \right|}^2} + \frac{1}{2}{\beta _2}{{\left| {{\psi _2}} \right|}^4} + {{\bar \kappa }_2}{{\left| {{\mathbf{\Pi }}{\psi _2}} \right|}^2}} \right]}d{\text{V}},
\end{equation}
the third ${\mathbb{F}_3}$ is
\begin{equation}
\label{Gibbs_energy3}
{\mathbb{F}_3} = \int\limits_{{\Omega _c}} {\left[ {{\alpha _3}{{\left| {{\psi _3}} \right|}^2} + \frac{1}{2}{\beta _3}{{\left| {{\psi _3}} \right|}^4} + {{\bar \kappa }_3}{{\left| {{\mathbf{\Pi }}{\psi _3}} \right|}^2}} \right]} d{\text{V}},
\end{equation}
and, finally, the interaction term $\mathbb{F}_{\operatorname{int}}$ is represented by the expression
\begin{equation}
\label{Gibbs_energy_int}
\begin{gathered}
  {\mathbb{F}_{\operatorname{int} }} =  - \int\limits_{{\Omega _c}} {\left[ {{\gamma _{12}}\left( {\psi _1^*{\psi _2} + {\psi _1}\psi _2^*} \right) + {\gamma _{13}}\left( {\psi _1^*{\psi _3} + {\psi _1}\psi _3^*} \right)} \right.}  \hfill \\
\qquad + \left. {{\gamma _{23}}\left( {\psi _2^*{\psi _3} + {\psi _2}\psi _3^*} \right)} \right]d{\text{V}} \hfill \\ 
\end{gathered} 
\end{equation}
with $\psi _{i} =\left|\psi _{i} \right|\exp \left(i\chi _{i} \right)$ being the three-component order parameter function and ${\mathbf{\Pi }} \equiv  - i\hbar \nabla  - \frac{{2e}}{c}{\mathbf{A}}$. Below we suppose the inter-component interaction constants $\gamma _{ij} $ are small enough, that is $\left|\psi _{i} \right|\left(r\right)\approx {\rm const}$ fulfilled. The 2D integrations in Eqs. (\ref{Gibbs_energy1})-(\ref{Gibbs_energy_int}) are carried out over the cross-section of the superconductor ($\Omega _{c} $) in the square-bracketed terms, and over the cross-sections of the superconductor and of the open-bracketed ones ($\Omega _{c} +\Omega _{o} $) in the last (magnetic) terms Eq. (\ref{Gibbs_energy}). The double-connectedness of the cylinder is accounted for by the condition 
\begin{equation}
\label{quantization_condition}
\oint\limits_\Gamma  {\nabla {\chi _i} \cdot d{\mathbf{l}}}  = 2\pi {N_i},
\end{equation}
where $\Gamma$ is an arbitrary closed contour that lies inside the wall of the cylinder and encircles the opening,$\chi _{i} $ are order parameter phases and $\; N_{i} =0,\pm 1,\pm 2,...$ are winding numbers. It should be emphasized that there are no a priori reasons for setting $N_{1} =N_{2} =N_{3} $. As in the case of fractional vortices in bulk multi-component superconductor s nontrivial topological states arise when$N_{i} \ne N_{j} $ at least. In the presence of inter-component coupling, they are of the soliton type.
Here we adopt the model proposed in Refs. \onlinecite{Tanaka1} and \onlinecite{Yerin2} which assumes the amplitudes of the order parameters $\left|\psi _{i} \right|$ to be  constant since the inter-component interaction is weak. This approximation leads for some set of parameters to the exactly solvable double sine-Gordon equation, which solution can certainly provide insight into the rich physics of topological objects related to the multicomponent superconducting order parameter. 

 In weak fields we set $\left|\psi _{i} \right|$ equal to the equilibrium values for an unperturbed three-component superconductor. The variation procedure for the phases of the order parameters yields two differential equations (see Appendix A):
\begin{equation}
\label{phi_eq}
\begin{gathered}
  \frac{{\xi _1^2}}{{{R^2}}}\frac{{{d^2}\phi }}{{d{\varphi ^2}}} - {\gamma _{12}}\left( {\frac{{\left| {{\psi _2}} \right|}}{{\left| {{\psi _1}} \right|}} + \frac{{\left| {{\psi _1}} \right|}}{{{\kappa _2}\left| {{\psi _2}} \right|}}} \right)\sin \phi  -  \hfill \\
  \frac{{{\gamma _{13}}\left| {{\psi _3}} \right|}}{{\left| {{\psi _1}} \right|}}\sin \theta  + \frac{{{\gamma _{23}}\left| {{\psi _3}} \right|}}{{{\kappa _2}\left| {{\psi _2}} \right|}}\sin \left( {\theta  - \phi } \right) = 0, \hfill \\ 
\end{gathered}
\end{equation}
\vspace{-\baselineskip}
\begin{equation}
\label{theta_eq}
\begin{gathered}
 \qquad \frac{{\xi _1^2}}{{{R^2}}}\frac{{{d^2}\theta }}{{d{\varphi ^2}}} - \frac{{{\gamma _{12}}\left| {{\psi _2}} \right|}}{{\left| {{\psi _1}} \right|}}\sin \phi  -  \hfill \\
  {\gamma _{13}}\left( {\frac{{\left| {{\psi _3}} \right|}}{{\left| {{\psi _1}} \right|}} + \frac{{\left| {{\psi _1}} \right|}}{{{\kappa _3}\left| {{\psi _3}} \right|}}} \right)\sin \theta  - \frac{{{\gamma _{23}}\left| {{\psi _2}} \right|}}{{{\kappa _3}\left| {{\psi _3}} \right|}}\sin \left( {\theta  - \phi } \right) = 0. \hfill \\ 
\end{gathered} 
\end{equation}

Here $\chi _{1} -\chi _{2} =\phi $ and $\chi _{1} -\chi _{3} =\theta $, ${\xi _1}$ is the coherence length for the first component in the absence of the inter-component interaction, $\kappa _{2} =\bar{\kappa }_{2} /\bar{\kappa }_{1} =D_{2} /D_{1} $, $\kappa _{3} =\bar{\kappa }_{3} /\bar{\kappa }_{1} =D_{3} /D_{1} $, where $D_{i} $ are the intra-component diffusion coefficients and $\bar{\kappa }_{1} =1$. 

The system of Eqs. (\ref{phi_eq}) and (\ref{theta_eq}) should be supplemented by the appropriate boundary conditions:
\begin{equation}
\label{bc_phi}
\phi \left( {2\pi } \right) - \phi \left( 0 \right) = 2\pi {n_2},{\text{ }}{\left. {\frac{{d\phi }}{{d\varphi }}} \right|_{\varphi  = 0}} = {\left. {\frac{{d\phi }}{{d\varphi }}} \right|_{\varphi  = 2\pi }},
\end{equation}
\vspace{-\baselineskip}
\begin{equation}
\label{bc_theta}
\theta \left( {2\pi } \right) - \theta \left( 0 \right) = 2\pi {n_3},{\text{ }}{\left. {\frac{{d\theta }}{{d\varphi }}} \right|_{\varphi  = 0}} = {\left. {\frac{{d\theta }}{{d\varphi }}} \right|_{\varphi  = 2\pi }},
\end{equation}
where we have introduced new winding numbers: $n_{2} =N_{1} -N_{2} $ and $n_{3} =N_{1} -N_{3} $.

\section{Analytical solutions for phase solitons}

\subsection{General expressions}

In general, the system of Eqs. (\ref{phi_eq}) and (\ref{theta_eq}) has no analytical solutions and a numerical analysis is necessary. But for a qualitative insight at first we investigate two ``classical'' special cases of BTRS, namely:$\gamma _{12} =1$, $\gamma _{13} =1$,$\gamma _{23} =-1$ and $\gamma _{12} =-1$, $\gamma _{13} =-1$ ,$\gamma _{23} =-1$ with coinciding moduli of the order parameters $\left|\psi _{1} \right|=\left|\psi _{2} \right|=\left|\psi _{3} \right|=\left|\psi \right|$, where analytical solutions can be obtained. We make a further simplification and assume that $\kappa _{2} \to 0$, i.e. the intraband diffusion coefficient $D_{2} \to 0$. Then in the first case the system of Eqs. \ref{phi_eq} and \ref{theta_eq} will be reduced to the form corresponding to the so-called double sine-Gordon equation, only
\begin{equation}
\label{theta_eq_new}
\frac{{{d^2}\theta }}{{d{\varphi ^2}}} - {K_3}\left( {\sin \theta  \pm \cos \frac{\theta }{2}} \right) = 0,
\end{equation}
\vspace{-\baselineskip}
\begin{equation}
\label{phi_eq_new}
\phi  =  \pm \frac{\pi }{2} + \frac{\theta }{2},
\end{equation}
and in the second case to
\begin{equation}
\label{theta_eq_new2}
\frac{{{d^2}\theta }}{{d{\varphi ^2}}} + {K_3}\left( {\sin \theta  \pm \sin \frac{\theta }{2}} \right) = 0,
\end{equation}
\vspace{-\baselineskip}
\begin{equation}
\label{phi_eq_new2}
\phi  =  \pm \pi  + \frac{\theta }{2},
\end{equation}
where $K_{3} =\frac{R^{2} }{\xi _{1}^{2} } \left(1+\frac{1}{\kappa _{3} } \right)$. Thereby the following simplified boundary conditions must be obeyed:
\begin{equation}
\label{bc_new}
  \theta \left( {2\pi } \right) - \theta \left( 0 \right) = 2\pi {n_3}, {\left. {\frac{{d\theta }}{{d\varphi }}} \right|_{\varphi  = 0}} = {\left. {\frac{{d\theta }}{{d\varphi }}} \right|_{\varphi  = 2\pi }}. 
\end{equation}

 Note that for odd $n_{3}$ the phase difference $\theta \left(2\pi \right)-\theta \left(0\right)=\pi n_{3} $. It means that a domain wall occurs at $\varphi =0$ for odd $n_{3} $. It costs no energy due to the assumption $\kappa _{2} =0$.
 
 Solving Eqs. \ref{theta_eq_new} and \ref{phi_eq_new} yields
\begin{equation}
\label{theta_sol_new1}
{\theta _{{n_3}}} =  \pm 4\arctan \left[ {\frac{{a_{{n_3}}^{\left( i \right)}{\text{sn}}\left( {{b_{{n_3}}}\left( {\varphi  - \pi } \right),{k_{{n_3}}}} \right) - 1}}{{a_{{n_3}}^{\left( i \right)}{\text{sn}}\left( {{b_{{n_3}}}\left( {\varphi  - \pi } \right),{k_{{n_3}}}} \right) + 1}}} \right],
\end{equation}
\vspace{-\baselineskip}
\begin{equation}
\label{phi_sol_new1}
{\phi _{{n_3}}} =  \pm \arctan \left[ {\frac{{2a_{{n_3}}^{\left( i \right)}{\text{sn}}\left( {{b_{{n_3}}}\left( {\varphi  - \pi } \right),{k_{{n_3}}}} \right)}}{{{{\left[ {a_{{n_3}}^{\left( i \right)}} \right]}^2}{\text{s}}{{\text{n}}^2}\left( {{b_{{n_3}}}\left( {\varphi  - \pi } \right),{k_{{n_3}}}} \right) - 1}}} \right],
\end{equation}
while the case with all repulsive inter-component interactions the solution of Eqs. (\ref{theta_eq_new2}) and (\ref{phi_eq_new2} gives 
\begin{equation}
\label{theta_sol_new2}
{\theta _{{n_3}}} =  \pm 4\arctan \left[ {a_{{n_3}}^{\left( i \right)}{\text{sn}}\left( {{b_{{n_3}}}\left( {\varphi  - \pi } \right),{k_{{n_3}}}} \right)} \right],
\end{equation}
\vspace{-\baselineskip}
\begin{equation}
\label{phi_sol_new2}
{\phi _{{n_3}}} =  \pm 2\arctan \left[ {a_{{n_3}}^{\left( i \right)}{\text{sn}}\left( {{b_{{n_3}}}\left( {\varphi  - \pi } \right),{k_{{n_3}}}} \right)} \right],
\end{equation}
where ${\rm sn}\left(z,\kappa \right)$ denotes the Jacobi elliptic sine function with the coefficients 
\begin{equation}
\begin{gathered}
  a_{{n_3}}^{\left( 1 \right)} = \sqrt { - \frac{{{C_{{n_3}}} - 3{K_3} + 2\sqrt {3K_3^2 - 2{C_{{n_3}}}{K_3}} }}{{{C_{{n_3}}} + 3{K_3}}}} , \hfill \\
  a_{{n_3}}^{\left( 2 \right)} = \sqrt { - \frac{{{C_{{n_3}}} + 3{K_3}}}{{{C_{{n_3}}} - 3{K_3} - 2\sqrt {3K_3^2 - 2{C_{{n_3}}}{K_3}} }}} , \hfill \\
  {b_{{n_3}}} = \sqrt { - \frac{{{C_{{n_3}}} - 3{K_3} - 2\sqrt {3K_3^2 - 2{C_{{n_3}}}{K_3}} }}{8}},  \hfill \\ 
\end{gathered}
\end{equation}
 and the modulus of the Jacobi elliptic function 
 \begin{equation}
 k_{n_{3} } =\sqrt{\frac{C_{n_{3} } -3K_{3} +2\sqrt{3K_{3}^{2} -2C_{n_{3} } K_{3} } }{C_{n_{3} } -3K_{3} -2\sqrt{3K_{3}^{2} -2C_{n_{3} } K_{3} } } }\hspace{0.1cm}.
 \end{equation}
 The set of constants $C_{n_{3} } \equiv {C}$ (see Appendix A) can be found from the appropriate transcendental equations in the case when only one of the intercomponent interactions is repulsive
\begin{equation}
\label{constant1_odd}
{\left( {a_{{n_3}}^{\left( i \right)}} \right)^2}{\text{s}}{{\text{n}}^2}\left( {{b_{{n_3}}}\pi ,{k_{{n_3}}}} \right) = 1, \quad \textrm{for odd} \quad n_3 =  \pm 1, \pm 3,...
\end{equation}
\vspace{-\baselineskip}
\begin{equation}
\label{constant1_even1}
2\left| {{n_3}} \right|\left( {{\text{K}} + \rm{i}{\text{K\ensuremath{'}}}} \right) = {b_{{n_3}}}\pi, \quad \textrm{for even} \quad n_3 =  \pm 2, \pm 6,...
\end{equation}
\vspace{-\baselineskip}
\begin{equation}
\label{constant1_even2}
2\left| {{n_3}} \right|\left( {{\text{K}} + \rm{i}{\text{K\ensuremath{'}}}} \right) + \rm{i}{\text{K\ensuremath{'}}} = {b_{{n_3}}}\pi, \quad \textrm{for even} \quad n_3 =  \pm 4, \pm 8,...
\end{equation}
 and also for the case of all repulsive inter-component interactions
\begin{equation}
\label{constant2_even1}
2\left| {{n_3}} \right|\left( {{\text{K}} + \rm{i}{\text{K\ensuremath{'}}}} \right) = {b_{{n_3}}}\pi,   \quad  \textrm{for odd} \quad {n_3} =  \pm 1, \pm 3,...
\end{equation}
\vspace{-\baselineskip}
\begin{equation}
\label{constant2_even2}
2\left| {{n_3}} \right|\left( {{\text{K}} + \rm{i}{\text{K\ensuremath{'}}}} \right) + \rm{i}{\text{K\ensuremath{'}}} = {b_{{n_3}}}\pi, \quad \textrm{for even} \quad {n_3} =  \pm 2, \pm 4,...
\end{equation}
where $\rm{i}=\sqrt { - 1} $ is the imaginary unit, ${\rm K}$ and ${\rm K}\ensuremath{'}$ are complete elliptic integrals of the first kind with the modulus $k_{n_{3} } $ and the complementary modulus $\sqrt{1-k_{n_{3} }^{2} } $ respectively.

Thus, there are two possible types of solitonic states in a three-component superconductor with a BTRS ground state differing in their winding numbers. Several examples of solitonic solutions with specific (certain) winding numbers are shown in the Appendix B.

\subsection{Stability, self-energy and Gibbs energy}

 In the limiting cases considered here, the expression for the Gibbs free energy of a three-component superconductor is given by 
\begin{equation}
\label{Gibbs_limiting}
G = {F_0} +  \frac{{\xi _1^2}}{{{R^2}}}\left| {{\alpha _1}} \right|{\left| \psi  \right|^2}V_s{\left( {{N_1} - \frac{{{k_3}}}{{1 + {k_3}}}{n_3} - \frac{\Phi }{{{\Phi _0}}}} \right)^2} + {F_{sol}},
\end{equation}
where $V_s=2\pi RLd$ (see Fig. \ref{cylinder}), $F_{0} $ is the free energy of the unperturbed superconducting cylinder and $F_{sol} $ is the contribution of the soliton self-energy, which is defined in the first case with one repulsive inter-component interaction by
\begin{equation} 
\label{Fsol1} 
\begin{gathered}
  {F_{sol}} = {\left| \psi  \right|^2}\frac{V_s}{{{K_3}}}{\int\limits_0^{2\pi } {\left( {\frac{{d{\theta _{{n_3}}}\left( \varphi  \right)}}{{d\varphi }}} \right)} ^2}d\varphi  +  \hfill \\
 2{\left| \psi  \right|^2}L\int\limits_0^{2\pi } {\left( {\frac{3}{2}  - \cos {\theta _{{n_3}}}\left( \varphi  \right) \pm 2\sin \frac{{{\theta _{{n_3}}}\left( \varphi  \right)}}{2}} \right)d\varphi } \hspace{0.1 cm},  \hfill \\ 
\end{gathered}  
\end{equation} 
and for the second case (all inter-component interactions are repulsive)
\begin{equation} 
\label{Fsol2} 
\begin{gathered}
  {F_{sol}} = {\left| \psi  \right|^2}\frac{V_s}{{{K_3}}}{\int\limits_0^{2\pi } {\left( {\frac{{d{\theta _{{n_3}}}\left( \varphi  \right)}}{{d\varphi }}} \right)} ^2}d\varphi  -  \hfill \\
  2{\left| \psi  \right|^2}L\int\limits_0^{2\pi } {\left( {\frac{3}{2} - \cos {\theta _{{n_3}}}\left( \varphi  \right) \pm 2\cos \frac{{{\theta _{{n_3}}}\left( \varphi  \right)}}{2}} \right)d\varphi }\hspace{0.1 cm}.  \hfill \\ 
\end{gathered}
\end{equation} 

The solutions of Eqs. (\ref{theta_sol_new1})-(\ref{phi_sol_new2}) considered as stationary points of the Gibbs free-energy functional  Eq. (\ref{Gibbs_limiting}), can correspond to either minima or saddle points. Taking into account that only stable phase configurations are physically meaningful, we have to turn to the sufficient conditions of a minimum, which requires also an analysis of the second variations of Eqs. (\ref{Fsol1}) and (\ref{Fsol2}). To this end we should treat the Sturm-Liouville problem for a three-component superconductor  with $\gamma _{12} =1$, $\gamma _{13} =1$ ,$\gamma _{23} =-1$
\begin{equation}
\label{stability1}
 - \frac{{{d^2}\eta }}{{d{\varphi ^2}}} + \left( {\cos {\theta _{{n_3}}}\left( \varphi  \right) \mp \frac{1}{2}\sin \frac{{{\theta _{{n_3}}}\left( \varphi  \right)}}{2}} \right)\eta  = \mu \eta ,
\end{equation}
and for $\gamma _{12} =-1$, $\gamma _{13} =-1$, $\gamma _{23} =-1$
\begin{equation}
 \label{stability2} 
-\frac{d^{2} \eta }{d\varphi ^{2} } -\left(\cos \theta _{n_{3} } \left(\varphi \right)\mp \frac{1}{2} \cos \frac{\theta _{n_{3} } \left(\varphi \right)}{2} \right)\eta =\mu \eta ,     
\end{equation} 
supplemented by the appropriate boundary conditions
\begin{equation}
 \label{bc_stability} 
\eta \left( 0 \right) = \eta \left( {2\pi } \right), {\left. {\frac{{d\eta }}{{d\varphi }}} \right|_{\varphi  = 0}} = {\left. {\frac{{d\eta }}{{d\varphi }}} \right|_{\varphi  = 2\pi }},    
\end{equation} 
where the $\theta _{n_{3} } \left(\varphi \right)$ are given by Eqs. (\ref{theta_sol_new1}) and (\ref{theta_sol_new2}).
\begin{figure}
\includegraphics[width=0.99\columnwidth]{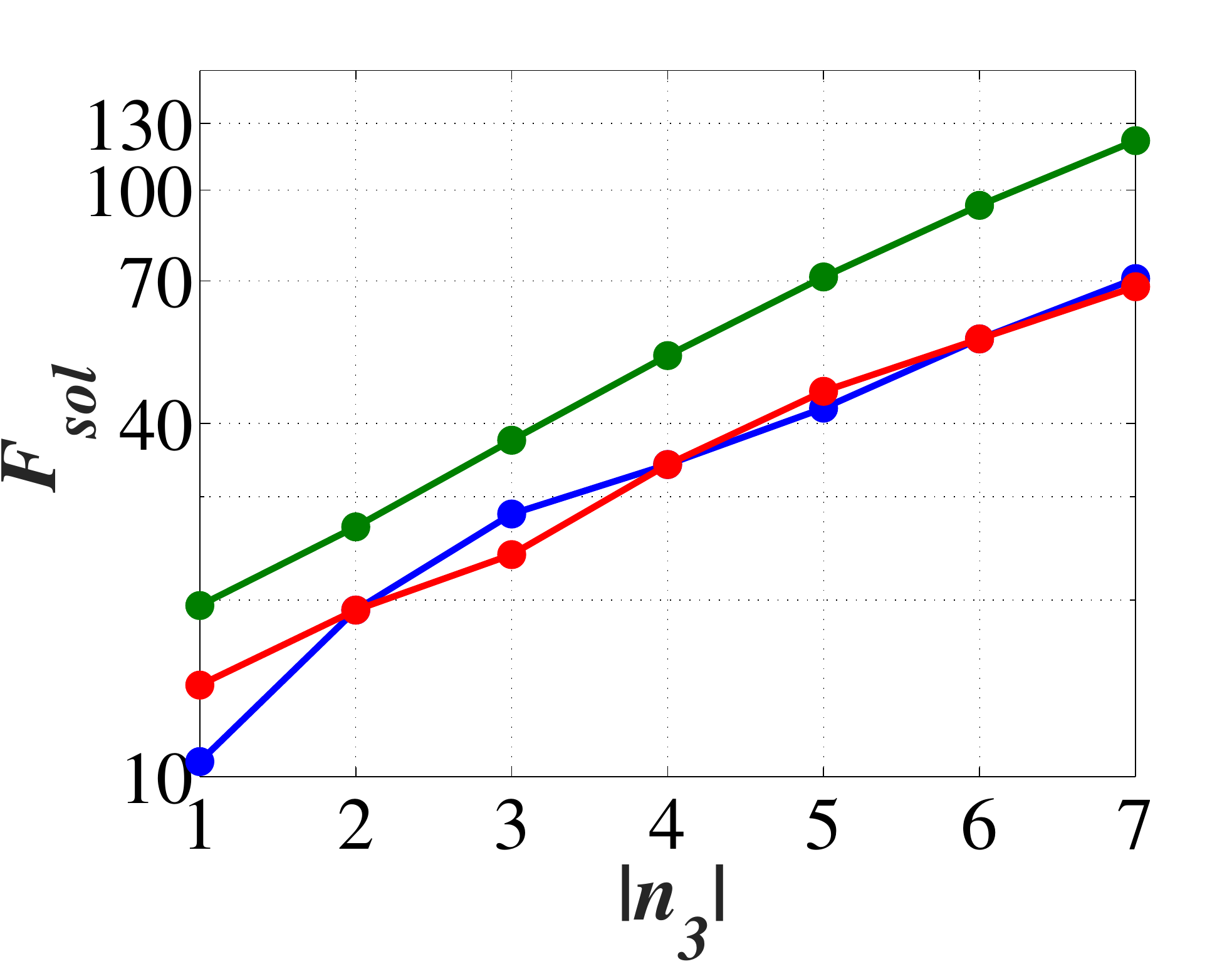}
\caption {Dependence of the soliton self-energy in units of $\frac{{\xi _1^2}}{{{R^2}}}\left| {{\alpha _1}} \right|{\left| \psi  \right|^2}V_s$ in a semi-logarithmic scale in a BTRS three-component superconductor with one repulsive inter-component interaction ($\gamma _{12} =1$, $\gamma _{13} =1$, $\gamma _{23} =-1$; blue and red lines respectively) and with all repulsive inter-component interactions ($\gamma _{12} =-1$, $\gamma _{13} =-1$, $\gamma _{23} =-1$; green line) on the winding number $n_{3} $.  Blue and red lines represent the self-energy for two types of phase solitons with $a^{\left(1\right)} $ coefficient and $a^{\left(2\right)} $ in Eq. (\ref{theta_sol_new1}) respectively for the case of a three-component superconductor  with $\gamma _{12} =1$, $\gamma _{13} =1$, $\gamma _{23} =-1$. The green line corresponds to the self-energy of phase solitons regardless of their type (regardless of coefficients of $a^{\left(i\right)} $  Eq. (\ref{theta_sol_new2}) for a three-component superconductor  with $\gamma _{12} =-1$, $\gamma _{13} =-1$, $\gamma _{23} =-1$ (see the text for details). The connecting broken lines are a guide for the eyes. Here $\kappa _{3} =0.5$ has been adopted.}
\label{self_energy}
\end{figure}
The lowest eigenvalues of the operators in Eqs. (\ref{stability1}) and (\ref{stability2}) can be found numerically and this way we obtained for both cases $\mu _{\min } =0$. This means that $\delta ^{2} F_{sol} \ge 0$ and the soliton states turn out to be indifferently stable states. The zero value of $\mu $ should be attributed to the existence of a zero-frequency ``rotational mode'' [Goldstone mode] (by analogy with the well-known translational Goldstone mode in quantum field theories \cite{Jackiw}) that restores the rotational symmetry broken by the formation of phase solitons.

Now, once the local stability of our soliton solutions is established, we can discuss their self-energies $F_{sol}$ expressed by Eqs. (\ref{Fsol1}) and (\ref{Fsol2}). By considering $\left|n_{3} \right|$ as a variable, we plot $F_{sol} $ as a function of the winding numbers (Fig. \ref{self_energy}).
First of all $F_{sol} \left(\left|n_{3} \right|\right)$ increases monotonously with an increase in $\left|n_{3} \right|$ as could be expected. Also, we can see that for a three-component superconductor  with one repulsive inter-component interaction and for a given {\it even} value of $n_{3} $ despite the presence of different types of phase solitons their self-energy remains the same while for an {\it odd} winding number, we observe an energy gap between their self-energies which however vanishes asymptotically with increasing $\left|n_{3} \right|$. Moreover, for a three-component superconductor with all repulsive inter-component interactions there is no difference in the self-energies of the two types of solitons for a given value of the winding number. 

Such an unusual behavior is caused by the structure of the solitonic solutions. As we can see from Eqs. (\ref{constant1_even1})-(\ref{constant2_even2}) for one repulsive inter-component interaction and for fixed \textit{even }$\left|n_{3} \right|$ as well as for all repulsive inter-component interactions and arbitrary $\left|n_{3} \right|$, the constants $C_{n_{3} } $ have the \textit{same} values for both types of phase solitons while for \textit{odd} values of the winding numbers Eq. \ref{constant1_odd} has \textit{distinct} roots (as a consequence distinct values of  $C_{n_{3} } $) for given $\left|n_{3} \right|$ due to different coefficients $a_{n_{3} }^{\left(i\right)} $. In turn, the set of constants $C_{n_{3} }$ can be considered as the value of the appropriate Lagrangians of the systems under consideration. Thus, despite the presence of different phase solitons both states are characterized by the same Lagrangians and as a result have equal self-energies (Fig. \ref{self_energy}).

In the same manner one can explain the decreasing difference between the various soliton self-energies for a three-component superconductor with one repulsive inter-component interaction. The numerical solution of Eq.  (\ref{constant1_odd}) demonstrates a decreasing difference between the values of $C_{n_{3} }$ for two types of solitons with an increasing winding number. This means that the energy difference starts to vanish, too (Fig. \ref{self_energy}). 

The damped oscillations of the soliton self-energies relative to each other in the case of a three-component superconductor with one repulsive inter-component interaction (see the blue and red curves in Fig. \ref{self_energy}) are connected with the distribution of the roots of Eq.  (\ref{constant1_odd}). According to the numerical solutions for winding numbers $\left|n_{3} \right|=4j+1$, where $j=0,1,2,...$, the  constants $C_{n_{3} } $ for the coefficient $a_{n_{3} }^{\left(1\right)} $ are always larger than for the coefficient $a_{n_{3} }^{\left(2\right)}$, while for $\left|n_{3} \right|=4j+3$ the constants $C_{n_{3} } $ for $a_{n_{3} }^{\left(1\right)}$ are always smaller than for $a_{n_{3} }^{\left(2\right)}$. In other words, in the case of odd values of $n_{3}$ for both types of phase solitons in a three-bans superconductor with one repulsive inter-component interaction peculiar alternation of the values of $C_{n_{3} }$ takes place: for $\left|n_{3} \right|=1$ and for $a_{1}^{\left(1\right)} $ the constant $C_{1}$ is larger than the analogical value of $C_{1}$ for $a_{1}^{\left(2\right)} $. Then for  $\left|n_{3} \right|=3$ and for $a_{3}^{\left(1\right)} $the value of $C_{3} $ is smaller than the counterpart $C_{3} $ for $a_{3}^{\left(2\right)} $ and so on. Bearing in mind that in fact the set of $C_{n_{3} }$ defines the self-energies of the phase solitons (see explanation given above), the above mentioned alternation explains the oscillatory behavior of the  two functions $F_{sol} \left(\left|n_{3} \right|\right)$ as shown in Fig. \ref{self_energy}.

We note that for the case of a three-component superconductor  with all inter-component interactions being repulsive, in the limit $\left|n_{3} \right|\to \infty $ or the same when ${k_{{n_3}}} \to 1$ we can use approximation for the elliptic sine in terms of hyperbolic tangent for Eq. (\ref{theta_sol_new2}) \cite{Abramowitz} and obtain phase soliton, which was found for a bulk BTRS three-component superconductor with infinite boundaries \cite{Lin2}.

\begin{figure}
\includegraphics[width=0.49\columnwidth]{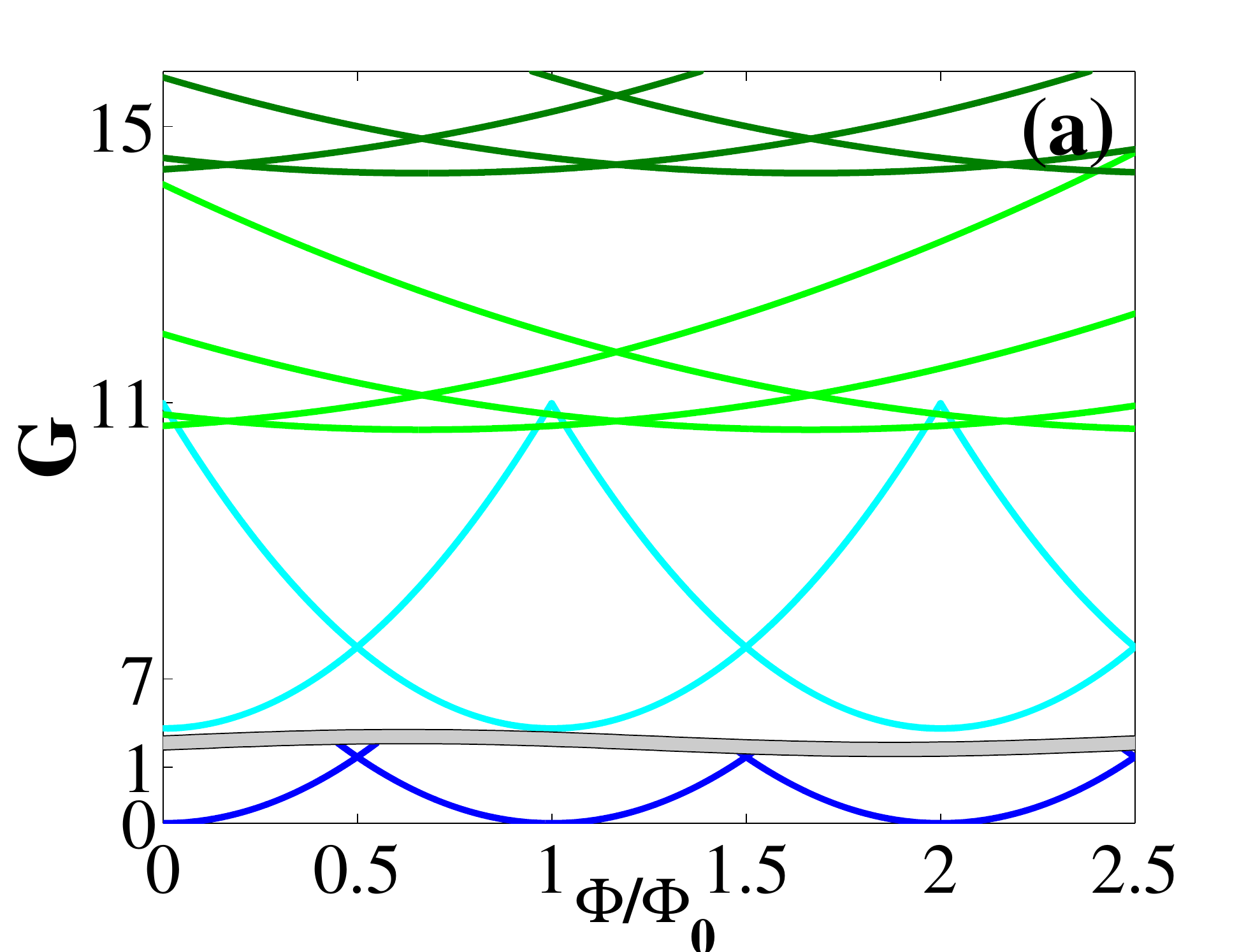}
\includegraphics[width=0.49\columnwidth]{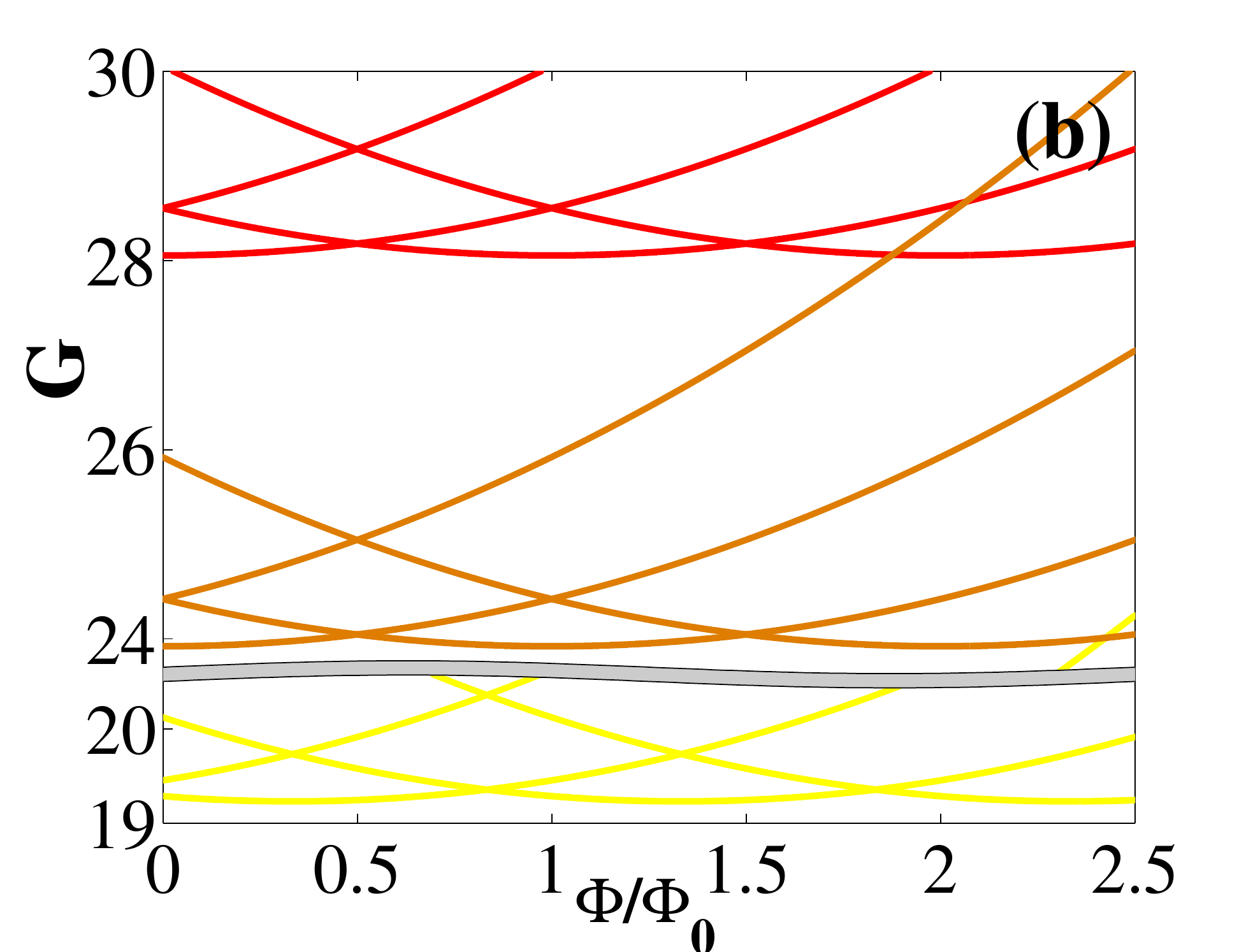}
\includegraphics[width=0.49\columnwidth]{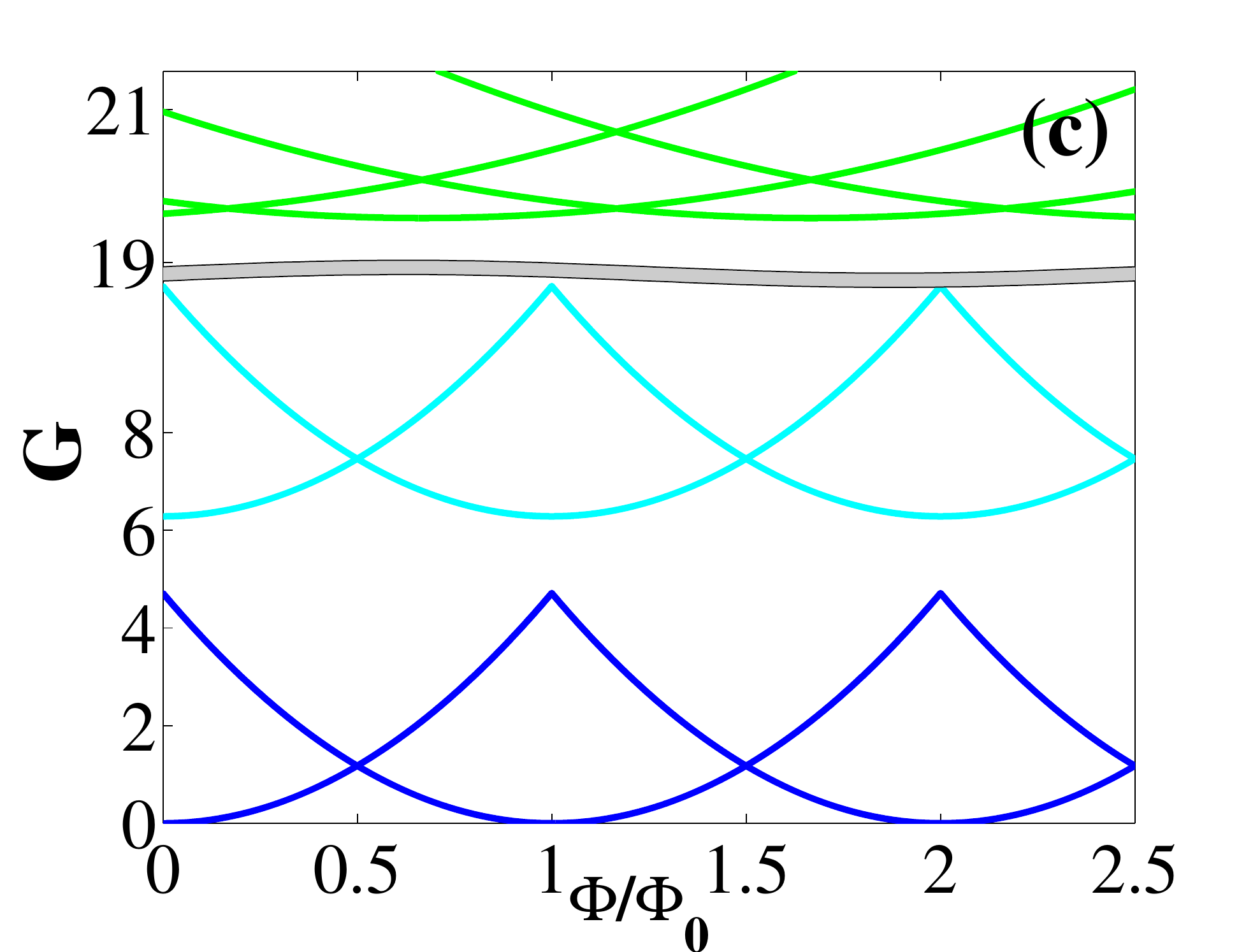}
\includegraphics[width=0.49\columnwidth]{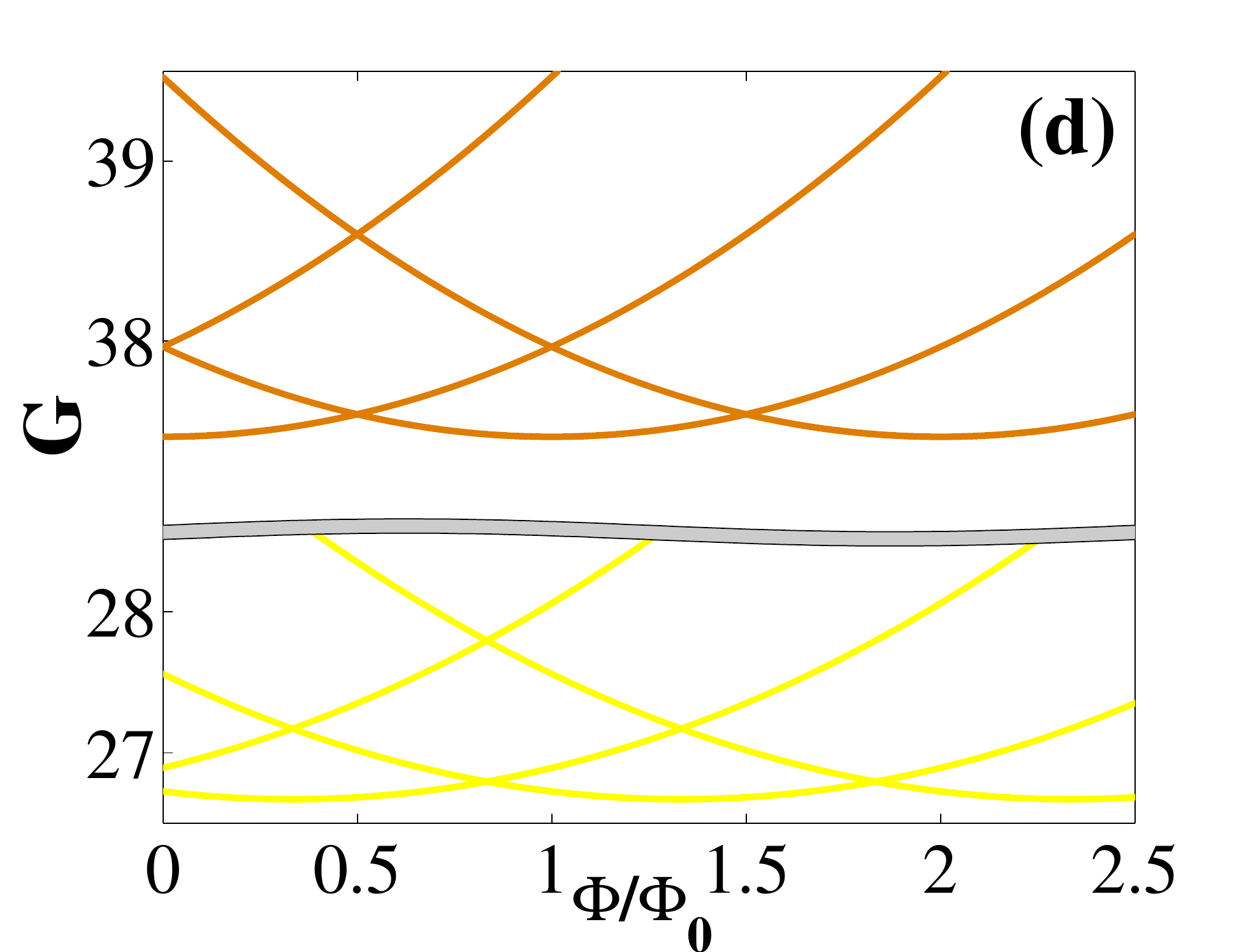}
\caption {Gibbs energy of homogeneous and inhomogeneous states in a BTRS three-component superconductor in units of $\frac{{\xi _1^2}}{{{R^2}}}\left| {{\alpha _1}} \right|{\left| \psi  \right|^2}V_s$ with $\gamma _{12} =1$, $\gamma _{13} =1$, $\gamma _{23} =-1$ (a, b) and $\gamma _{12} =-1$, $\gamma _{13} =-1$, $\gamma _{23} =-1$ (c, d). Blue lines corresponds to the energy of a homogeneous BTRS state with $\phi =\pm \pi /3$, $\theta =\mp \pi /3$ (a) or  $\phi =\pm 2\pi /3$, $\theta =\mp 2\pi /3$ (b). Cyan lines corresponds to homogeneous non-BTRS states with $\phi =0$, $\theta =\pi $ (a) and $\phi =\pi $, $\theta =\pi $ (b).  Other color curves illustrate the Gibbs free energy of topological defects in the form of phase solitons for a given set of winding numbers ($N_{1}, n_{3}$), where green and dark green lines represent solitonic solutions with $\left(N_{1} ,1\right)$ (a, c), yellow lines - $\left(N_{1} , 2\right)$ (b, d), orange and red lines - $\left(N_{1} ,3\right)$ (b, d). Here $\kappa _{3} =0.5$ has been adopted.}
\label{Gibbs_energy_analytical}
\end{figure}

Substituting Eqs. (\ref{theta_sol_new1}) and (\ref{theta_sol_new2}) into Eqs. (\ref{Fsol1}) and (\ref{Fsol2}) after a long analytical but straightforward integration we arrive at the total Gibbs free energy of a three-component superconductor (Fig. \ref{Gibbs_energy_analytical}).

As in the case of a two-band superconductor we can conclude that inhomogeneous states for a three-component superconductor cannot be the ground state of these systems, at least for this limiting case considered here.

\section{Numerical solutions for phase solitons}

\begin{figure}
\includegraphics[width=0.49\columnwidth]{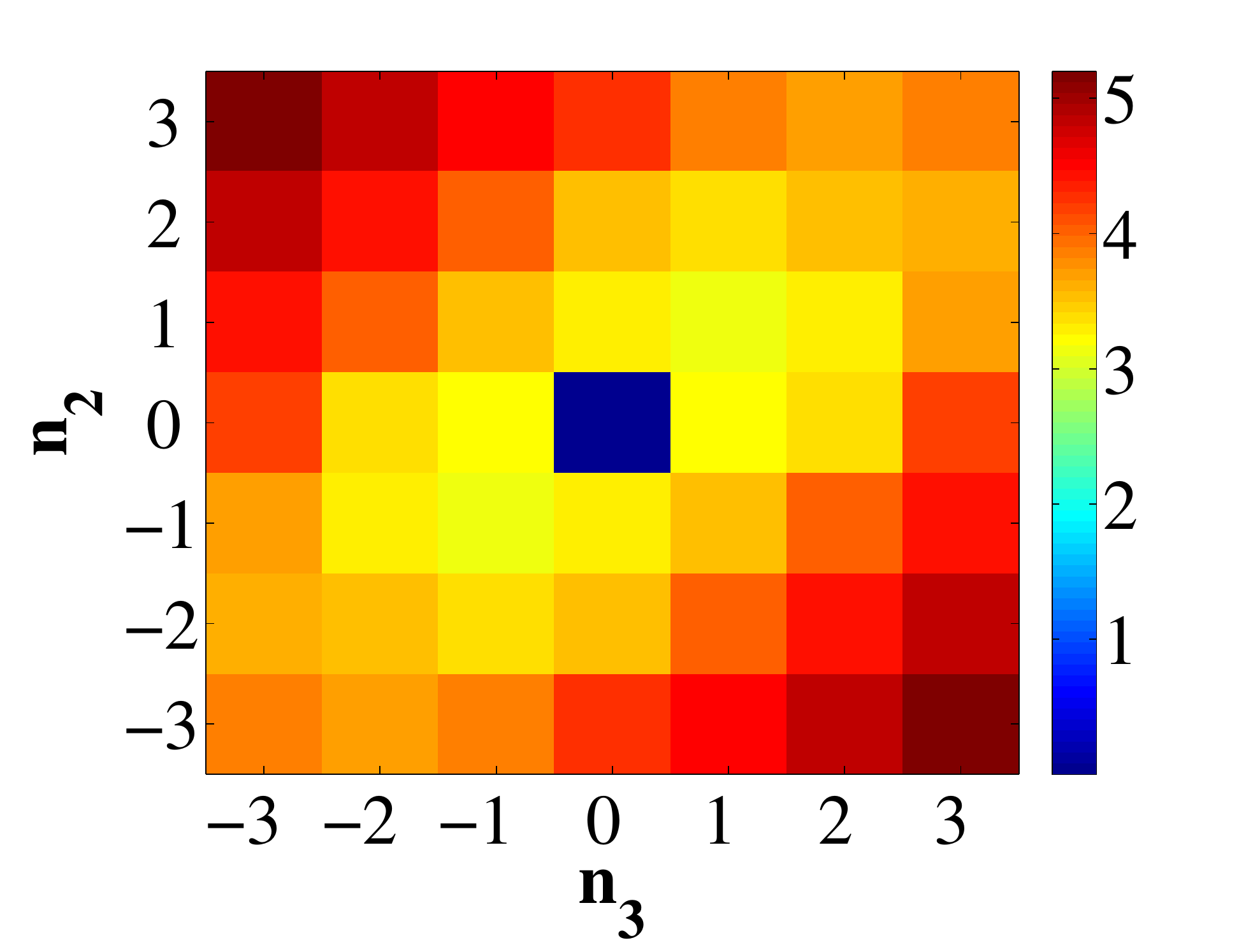}
\includegraphics[width=0.49\columnwidth]{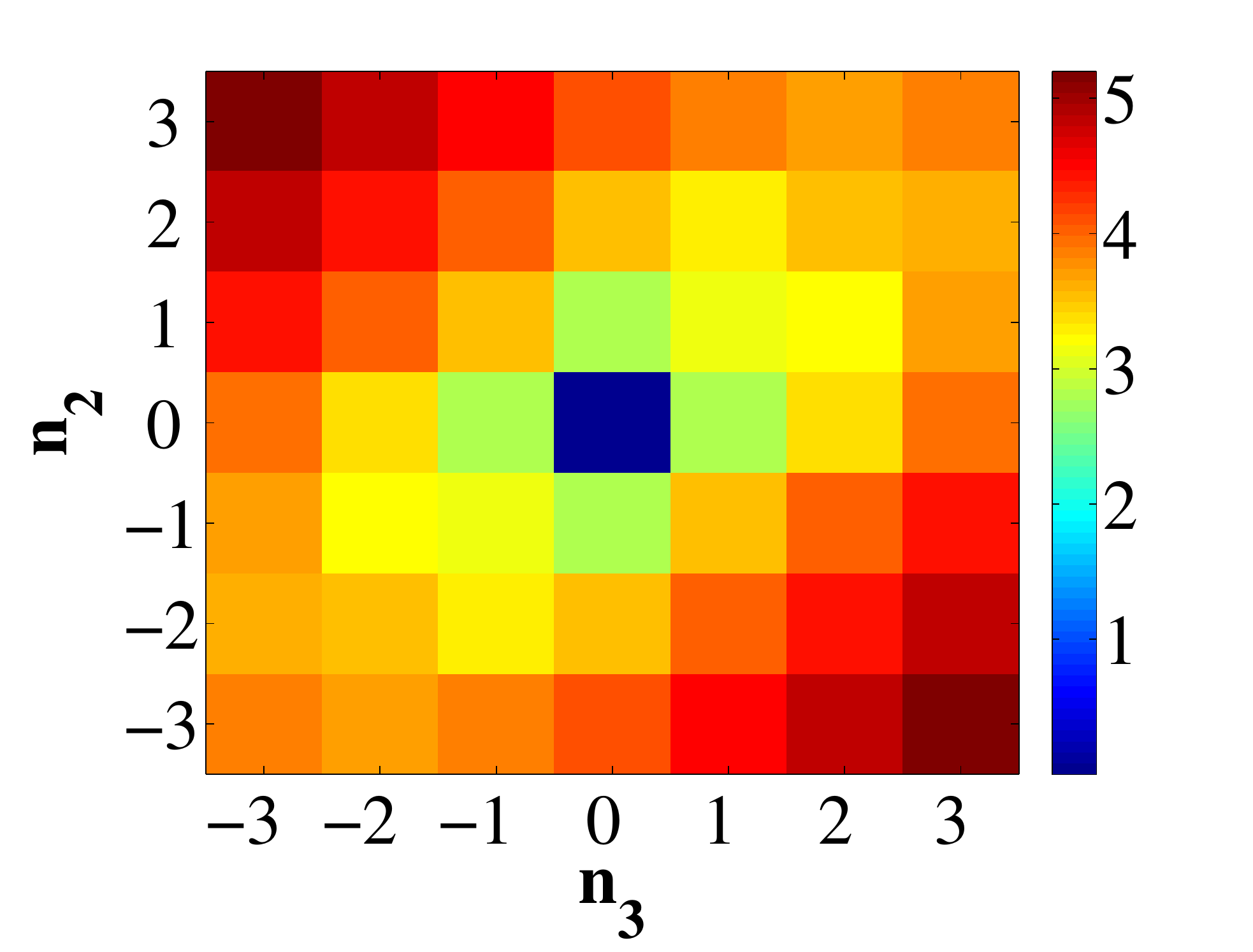}
\caption {Self-energy in units of $\frac{{\xi _1^2}}{{{R^2}}}\left| {{\alpha _1}} \right|{\left| \psi  \right|^2}V_s$ of two types of phase solitons in a BTRS three-component superconductor  with one repulsive inter-component interaction vs. different winding numbers plotted in a logarithmic scale for $\kappa _{2} =4$ and $\kappa _{3} =2$.}
\label{Fsol_numerical}
\end{figure}
Now we proceed to the numerical solution of Eqs. (\ref{phi_eq}) and (\ref{theta_eq}) with the boundary conditions Eqs. (\ref{bc_phi}) and (\ref{bc_theta}) for the set of parameters treated earlier, namely $\gamma _{12} =1$, $\gamma _{13} =1$ and $\gamma _{23} =-1$ at $\left|\psi _{1} \right|=\left|\psi _{2} \right|=\left|\psi _{3} \right|=\left|\psi \right|$. Here we will assume arbitrary ratios of the involved effective masses $\kappa _{2}$ and $\kappa _{3} $. From the analysis of the numerical solutions we obtain the dependence of the soliton self-energies $F_{sol}$ as a function of the winding numbers $n_{2} $ and $n_{3} $ for a BTRS three-component superconductor with one repulsive inter-component interaction (Fig.  \ref{Fsol_numerical}).

\begin{figure}
\includegraphics[width=0.99\columnwidth]{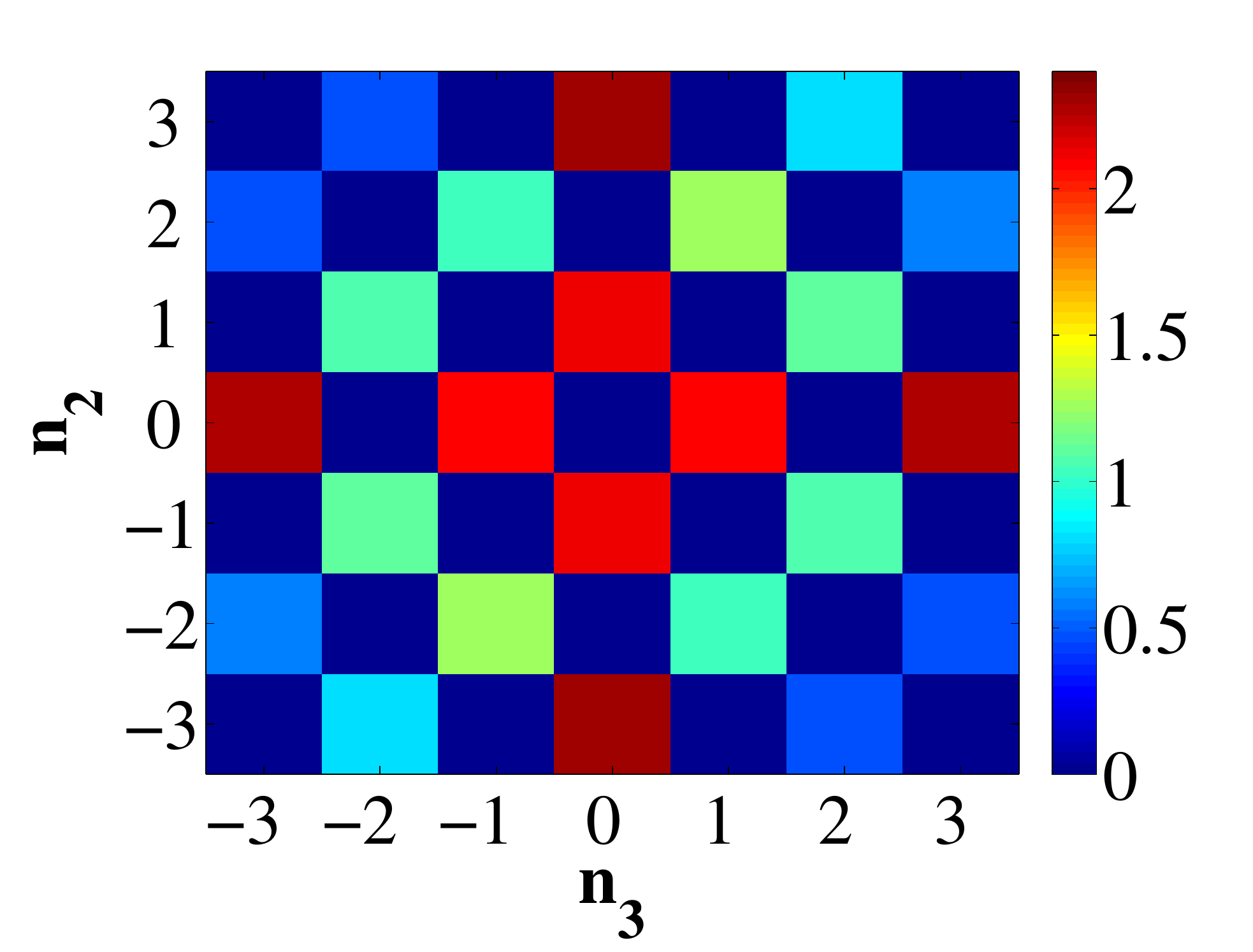}
\caption {The difference in the self-energy $F_{sol} $ in units of $\frac{{\xi _1^2}}{{{R^2}}}\left| {{\alpha _1}} \right|{\left| \psi  \right|^2}V_s$ of the two types solitonic states (see Fig. \ref{Fsol_numerical}) for given values of $n_{2} $ and $n_{3} $. For more clarity we plot the cubic root of the difference.}
\label{diff_Fsol_numerical}
\end{figure}

The analysis of $F_{sol}$ clearly demonstrates that the soliton self-energy is an even function with respect to the winding numbers, i.e. $F_{sol} \left(n_{2}, n_{3} \right)=F_{sol} \left(-n_{2}, -n_{3} \right)$ as it should be. Also, we can see that the self-energy of the solitons increases monotonically with an increase of  $n_{2}$ and $n_{3} $. Moreover, if the relation $\left|n_{2} -n_{3} \right|{\rm =0\; mod\; 2\; }$ is obeyed, then for both soliton solutions their self-energies do coincide (Fig. \ref{diff_Fsol_numerical}). 

\begin{figure}
\includegraphics[width=0.99\columnwidth]{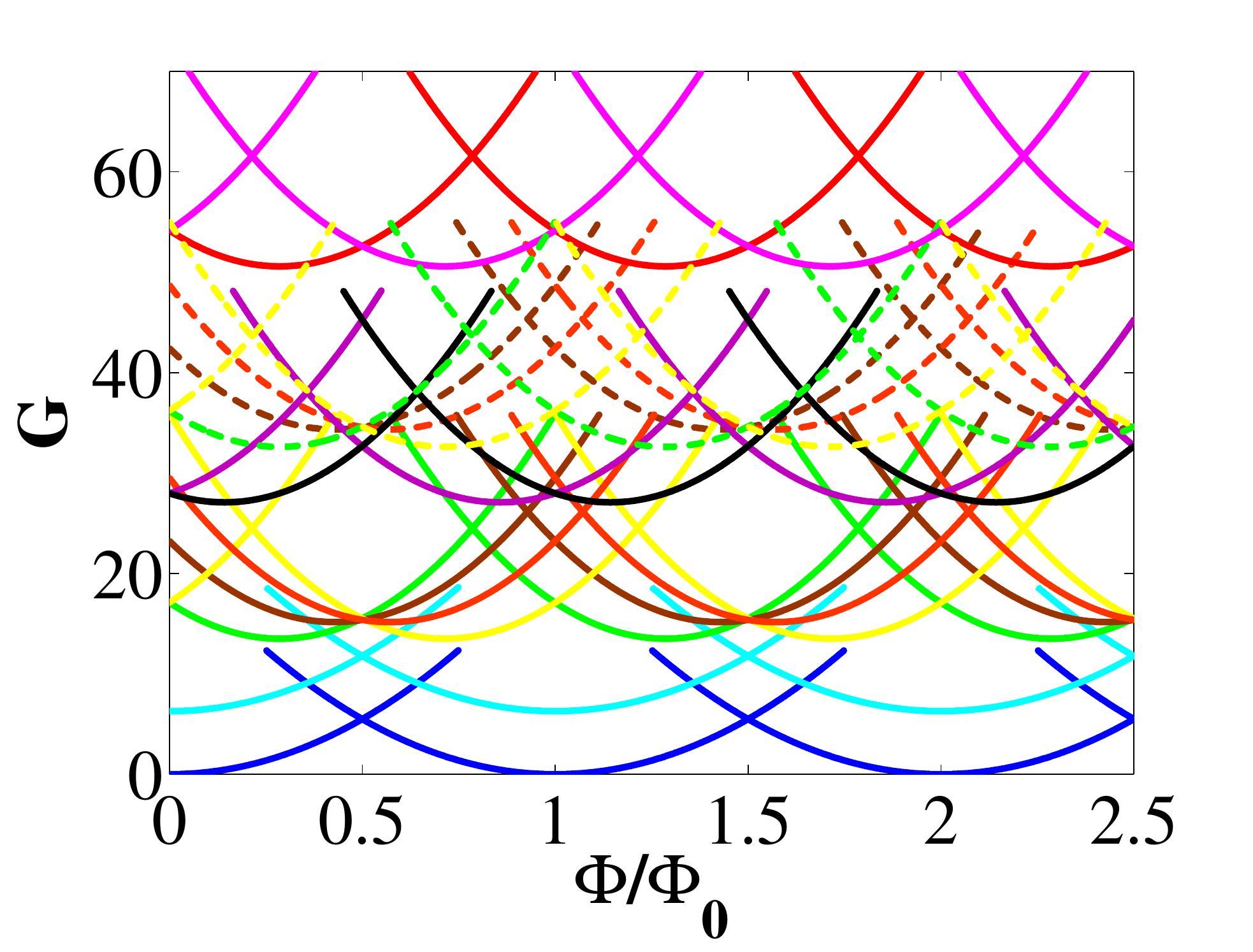}
\caption {Gibbs free energy of a three-component superconductor in units of $\frac{{\xi _1^2}}{{{R^2}}}\left| {{\alpha _1}} \right|{\left| \psi  \right|^2}V_s$ with one repulsive inter-component interaction  for homogeneous BTRS ground state (blue), homogeneous non-BTRS state (cyan) and different topological states $\left(N_{1}, n_{2}, n_{3} \right)$ for typical ratios of the effective masses $\kappa _{2} =4$, $\kappa _{3} =2$. Here green (solid and dashed) lines denote phase solitons with $\left(N_{1}, 0, -1\right)$, yellow (solid and dashed) - $\left(N_{1}, 0, 1\right)$, orange (solid and dashed) - $\left(N_{1}, -1, 0\right)$,  brown (solid and dashed)  - $\left(N_{1}, 1, 0\right)$, violet - $\left(N_{1}, -1, -1\right)$, black - $\left(N_{1}, 1, 1\right)$, red - $\left(N_{1}, -1, 1\right)$ and magenta  - $\left(N_{1}, 1, -1\right)$.}
\label{Gibbs_energy_numerical}
\end{figure}

Figure \ref{Gibbs_energy_numerical} focuses on the Gibbs free energy of different topological states $\left(N_{1}, n_{2}, n_{3} \right)$, where $\left|n_{2} \right|=0,1$ and $\left|n_{3} \right|=0,1$. The minima of the Gibbs free energy of solitonic states represent the soliton self-energy and occur when the self-induced flux compensates the external flux. 

Further numerical investigations reveal that for a three-component superconductor  withal inter-component interactions being repulsive, the key features of phase solitons, which were established in a particular case where analytical solutions are amendable, remain the same also for arbitrary values of $\kappa _{2}$ and $\kappa _{3}$.

Thus, as in the case of the analytical approach numerical calculations of the Gibbs free energy of a doubly-connected system display that the exact location of the phase solitons on the energetic scale of a BTRS three-component superconductor is always higher than the homogeneous BTRS and TRS states. This means that for a nonadiabatic, fast switched on magnetic flux, the mesoscopic thin rings or tubes made from three-component superconductors can be excited upon the BTRS ground state and thereby we can induce phase solitons passing another metastable TRS state. 

\section{Discussion}

From an experimental point of view transitions from the BTRS state to a TRS and then to phase solitons can be identified via the magnetic response. These transitions become visible by the presence of appropriate jumps on the current-magnetic flux dependencies. It should be mentioned that earlier such leaps were attributed to the BTRS-TRS transitions and were considered as a hall mark for the identification of BTRS multiband superconductivity. If however our system under consideration can be flipped also into states with phase solitons, then the dependencies of the current on the magnetic flux will be more complicated and it can acquire a significantly larger number of additional jumps in comparison with the same dependencies, when only one switch between homogeneous BTRS to TRS states and vice versa is allowed. In the former case we admit that the analysis of the response predicted by our proposed experiment for the detection of BTRS multiband superconductivity will meet some difficulties connected with the large zoo of phase soliton states. But in view of the fact that the probability of a transition is proportional to the Boltzmann factor $\exp \left( { - \frac{{\delta F}}{T}} \right)$, where $\delta F$ is the energy difference between the ground and the excited states for a given value of the magnetic flux, the generation of phase solitons is certainly not the dominant process.

It means that during experimental measurements BTRS-TRS transitions have a substantially higher probability than transitions from the BTRS ground state to states with phase solitons. Hence, the specific magnetic response of multiband superconductors remains nevertheless valid as a convenient tool for the detection of BTRS.

Despite the obvious limitations of the phenomenological Ginzburg-Landau theory one can perform rough estimates of the energy difference between solitonic states and the BTRS ground state in our three-component superconductor. The measure unit that is used here can be rewritten in terms of the magnetic inductance of the cylinder under consideration (or its self-inductance) $ \mathcal{L}_m= \frac{{4{\pi ^2}{R^2}}}{L}$, its radius (see Fig. \ref{cylinder}) and the London penetration depth $\lambda$: 
\begin{equation}
\label{num_unit}
\frac{{\xi _1^2}}{{{R^2}}}\left| {{\alpha _1}} \right|{\left| \psi  \right|^2}V_s =\frac{{dR}}{{2{\lambda ^2}}}\frac{{\Phi _0^2}}{{2{\mathcal{L}_m}}},
\end{equation}
where we adopt for  $\lambda$ the standard Ginzburg-Landau formula with the largest value of the order parameter modulus as an upper bound in Eq. (\ref{num_unit}) for the sake of simplicity and clarity \cite{Gennes, note2} and ${\Phi _0} = \frac{{\pi \hbar c}}{e}$ is the magnetic flux quantum. Based on the characteristics of a system that was used for the experimental detection of phase solitons in a two-component superconductor in Ref. \onlinecite{Bluhm}, namely $R= \SI{2}{\micro\metre}$ and for the certain close to $T_c$ temperature $\lambda  \approx  \SI{1}{\micro\metre}$ (see Fig. 2a in Ref. \onlinecite{Bluhm}) one can evaluate the numerical value of the introduced energetical unit $\frac{{\xi _1^2}}{{{R^2}}}\left| {{\alpha _1}} \right|{\left| \psi  \right|^2}V_s \approx 0.6{\text{ meV}}$.
As one can see from Figs. \ref{Gibbs_energy_analytical} and \ref{Gibbs_energy_numerical} the distance between the BTRS and the lowest solitonic states is about 10-20 $\frac{{\xi _1^2}}{{{R^2}}}\left| {{\alpha _1}} \right|{\left| \psi  \right|^2}V_s$  that in turn gives 6-12 meV. In the case $\text{Ba}_{\text{0.6}}\text{K}_{\text{0.4}}\text{Fe}_{\text{2}}\text{As}_{\text{2}}$ this is equal to approximately $\sim 0.5 - 1$ of the largest energy gap in this compound \cite{Ding} considered as an upper bound.

The possible strategy for the experimental detection of phase solitons is the application of a rapidly increasing magnetic field to the superconducting cylinder to induce transitions from the BTRS state directly to metastable solitonic states bypassing the intermediate TRS state. In other words, following the similar spirit of the Kibble-Zurek scenario of the generation of topological defects in superconductors \cite{Shapiro}, we can use a magnetic field as the driving tool that can flip the system from the ground state to an excited state with phase solitons. Since the energy of phase-inhomogeneous solutions is higher than those the BTRS and TRS states, one can in principle observe additional jumps on the current magnetic flux dependencies due to relaxation processes from higher energetic levels (soliton states) to the ground state via intermediate metastable TRS states. To illustrate the dynamic response and to characterize relaxation processes, a special analysis within time-dependent Ginzburg-Landau approach outside of the scope of the present paper is required and therefore left for future studies.

Also, we would like to remark that in the real experimental situation sketched in Ref. \onlinecite{Yerin2, Yerin3} and in Figure \ref{cylinder} the phase solitons are created strictly speaking dynamically, i.e. their creation energy includes also a final kinetic energy contribution ignored here. It will be considered in a forthcoming time-dependent Ginzburg-Landau approach \cite{Yerin4} together with other dynamical effects beyond the scope of the present initial static study. 

Finally, it is worth note that we exclude from the consideration the emergence of Fulde-Ferrell-Larkin-Ovchinnikov (FFLO) states in a three-component superconductor that can contribute to the diversity of topological states in such a system and  unambiguously complicate the exact location and further identification of phase solitons on the energy scale. Such a problem was not fully elucidated even for the case of a two-band superconductor \cite{Machida1, Ptok, Machida2}, where we admit coexistence and interplay of FFLO state with phase solitons, especially in the presence of interband scattering. We plan to resolve this issue and rank them on a energy scale in the near future.

\section{Conclusions}

To summarize, we have shown that three-component superconductor phase solitons on a cylinder may exist. These phase solitons are thermodynamically metastable. We have shown that the total energy of phase-inhomogeneous solutions is higher than that of homogeneous BTRS and TRS states. The Gibbs free energy of a three-component superconductor  increases monotonically with an increase of the winding numbers $\left|n_{2} \right|$ and $\left|n_{3} \right|$. For the case of a BTRS three-component superconductor  with one repulsive inter-component interaction two types of solitons for a given even numbered difference $\left|n_{2} -n_{3} \right|$ have equal self-energies, while for an odd numbered difference of winding numbers there is an energy gap which rapidly decreases with an increase of $\left|n_{2} \right|$ and $\left|n_{3} \right|$. For a BTRS three-component superconductor  with all inter-component interactions being repulsive, the self-energy of both types of phase solitons do coincide for all values of the winding numbers $\left|n_{2} \right|$ and $\left|n_{3} \right|$.

Finally, we draw attention to our suggestion that the very existence of phase solitons under consideration to the best of our knowledge may occur in microscopically inhomogeneous systems, only (probably beyond a critical threshold) described by the standard GL theory. In this context the occurrence of novel topological solutions detected experimentally would provide \textit{a posteriori} a justification for the application of specific phenomenologically introduced multi-component functionals doubted recently \cite{Ichioka1} based on global symmetry arguments for second order phase transitions, only. In other words, the microscopic inhomogeneity given for instance by separated Fermi surface sheets adds effective different quantum numbers to various groups (bands) of electrons as compared to the simple picture of a single-band isotropic superconductor with a single phase, in particular. It is convenient to describe mesoscopic \textit{inhomogeneous} states semi-quantitatively within a generalized quasi-GL approach adopted here, which would be very difficult within any microscopic approach. The full quantitative picture of phase solitonic states can be provided within the microscopic theory based on the generalization of BCS theory for three bands (see e.g. Ref \onlinecite{Drechsler1}. In this case it is possible to specify the borders of applicability of the Ginzburg-Landau theory that is exploited here. It needs a separate investigation, and noteworthy even for a two-band superconductor this subject is controversially discussed up to now \cite{Babaev_comment, Schmalian_reply}. To this end our work can be considered as a first phenomenological approximation to shed light on the emergence and the behavior of phase solitons in three-component (three-band) superconductors and superfluids.

The occurrence of novel topological solutions seems to be more important than the formal occurrence of new length scales due to a higher order expansion \cite{Komendova} with a somewhat unclear physical meaning.  In our opinion the account of higher order terms ignored here causes slight quantitative changes of the solitonic shapes, only.

Based on two supporting features (frustration and mesoscopic inhomogeneity) we provide a reasonable scenario for the observation of phase solitons in three-component superconductors with repulsive inter-component interaction and a \textit{posterio} justification of the employed effective GL functional in view of the observed experimental features \cite{Bluhm, Tanaka2}.

\begin{acknowledgments}

We thank D. Efremov, E. Babaev, J. van den Brink, B. Malomed, H.-H. Klauss and Y. Ovchinnikov for valuable, critical and constructive discussions as well as stimulating interest. Y.Y. acknowledges a financial support by the CarESS project and the Graduierten Kolleg of TU Dresden. Y.Y. thanks the IFW Dresden for hospitality, where parts of the present work were performed.

\end{acknowledgments}	

\begin{widetext}
\appendix
\section{Ginzburg-Landau equations for phase solitons and the first integrals}

We describe the order parameters in a three-band superconductor as  ${\psi _i} = \left| {{\psi _i}} \right|{e^{i{\chi _i}}}$, where  $\left| {{\psi _i}} \right|$ are order parameters moduli and  ${\chi _i}$  are their phases. A variation procedure with respect to ${\chi _i}$ yields
\begin{equation}
\label{Eq_GL1}
\frac{{{\hbar ^2}{{\left| {{\psi _1}} \right|}^2}}}{{{R^2}{m_1}}}\frac{{{d^2}{\chi _1}}}{{d{\varphi ^2}}} - {\gamma _{12}}\left| {{\psi _1}} \right|\left| {{\psi _2}} \right|\sin \left( {{\chi _1} - {\chi _2}} \right) - {\gamma _{13}}\left| {{\psi _1}} \right|\left| {{\psi _3}} \right|\sin \left( {{\chi _1} - {\chi _3}} \right) = 0,
\end{equation}
\vspace{-\baselineskip}
\begin{equation}
\label{Eq_GL2}
\frac{{{\hbar ^2}{{\left| {{\psi _2}} \right|}^2}}}{{{R^2}{m_2}}}\frac{{{d^2}{\chi _2}}}{{d{\varphi ^2}}} + {\gamma _{12}}\left| {{\psi _1}} \right|\left| {{\psi _2}} \right|\sin \left( {{\chi _1} - {\chi _2}} \right) - {\gamma _{23}}\left| {{\psi _2}} \right|\left| {{\psi _3}} \right|\sin \left( {{\chi _2} - {\chi _3}} \right) = 0,
\end{equation}
\vspace{-\baselineskip}
\begin{equation}
\label{Eq_GL3}
\frac{{{\hbar ^2}{{\left| {{\psi _3}} \right|}^2}}}{{{R^2}{m_3}}}\frac{{{d^2}{\chi _3}}}{{d{\varphi ^2}}} + {\gamma _{13}}\left| {{\psi _1}} \right|\left| {{\psi _3}} \right|\sin \left( {{\chi _1} - {\chi _3}} \right) + {\gamma _{23}}\left| {{\psi _2}} \right|\left| {{\psi _3}} \right|\sin \left( {{\chi _2} - {\chi _3}} \right) = 0.
\end{equation}

The lack of  $\alpha$ and $\beta$ in Eqs. (\ref{Eq_GL1})-(\ref{Eq_GL3}) is due to the fact these coefficients contribute to the Ginzburg-Landau functional with the absolute values of the order parameters (see Eqs. (\ref{Gibbs_energy1})-(\ref{Gibbs_energy_int})) and have no dependence on their phases in comparison with the kinetic (gradient) and interband interaction terms.

Dividing each of the Eqs. (\ref{Eq_GL1})-(\ref{Eq_GL3}) by the appropriate value of ${\left| {{\psi _i}} \right|^2}$  and then subtracting from the first equations the other two ones, we get the system of Eqs. (\ref{Eq_GL1_new}) and (\ref{Eq_GL2_new}) 
\begin{equation}
\label{Eq_GL1_new}
\frac{{\xi _1^2}}{{{R^2}}}\frac{{{d^2}\phi }}{{d{\varphi ^2}}} - {\gamma _{12}}\left( {\frac{{\left| {{\psi _2}} \right|}}{{\left| {{\psi _1}} \right|}} + \frac{{\left| {{\psi _1}} \right|}}{{{k_2}\left| {{\psi _2}} \right|}}} \right)\sin \phi  - \frac{{{\gamma _{13}}\left| {{\psi _3}} \right|}}{{\left| {{\psi _1}} \right|}}\sin \theta  + \frac{{{\gamma _{23}}\left| {{\psi _3}} \right|}}{{{k_2}\left| {{\psi _2}} \right|}}\sin \left( {\theta  - \phi } \right) = 0,
\end{equation}
\vspace{-\baselineskip}
\begin{equation}
\label{Eq_GL2_new}
\frac{{\xi _1^2}}{{{R^2}}}\frac{{{d^2}\theta }}{{d{\varphi ^2}}} - \frac{{{\gamma _{12}}\left| {{\psi _2}} \right|}}{{\left| {{\psi _1}} \right|}}\sin \phi  - {\gamma _{13}}\left( {\frac{{\left| {{\psi _3}} \right|}}{{\left| {{\psi _1}} \right|}} + \frac{{\left| {{\psi _1}} \right|}}{{{k_3}\left| {{\psi _3}} \right|}}} \right)\sin \theta  - \frac{{{\gamma _{23}}\left| {{\psi _2}} \right|}}{{{k_3}\left| {{\psi _3}} \right|}}\sin \left( {\theta  - \phi } \right) = 0.
\end{equation}
Here we have introduced the new phase variables ${\chi _1} - {\chi _2} = \phi$  and ${\chi _1} - {\chi _3} = \theta$  and the parameters ${k_2} = {m_1}/{m_2}$,  ${k_3} = {m_1}/{m_3}$.

The first integrals of the Eqs. (\ref{theta_eq_new}) and (\ref{theta_eq_new2}) in the main paper are
\begin{equation}
\label{FI_1}
\frac{1}{2}{\left( {\frac{{d\theta }}{{d\varphi }}} \right)^2} = C - {K_3}\left( {\cos \theta  \pm 2\sin \frac{\theta }{2}} \right),
\end{equation}
\vspace{-\baselineskip}
\begin{equation}
\label{FI_2}
\frac{1}{2}{\left( {\frac{{d\theta }}{{d\varphi }}} \right)^2} = C + {K_3}\left( {\cos \theta  \pm 2\cos \frac{\theta }{2}} \right)
\end{equation}
We should note that for both signs in the Eqs. (\ref{FI_1}) and (\ref{FI_2}), the  inequality $ - 3{K_3} \leqslant C <  + \infty $  must be satisfied.
\begin{figure}
\includegraphics[width=0.49\columnwidth]{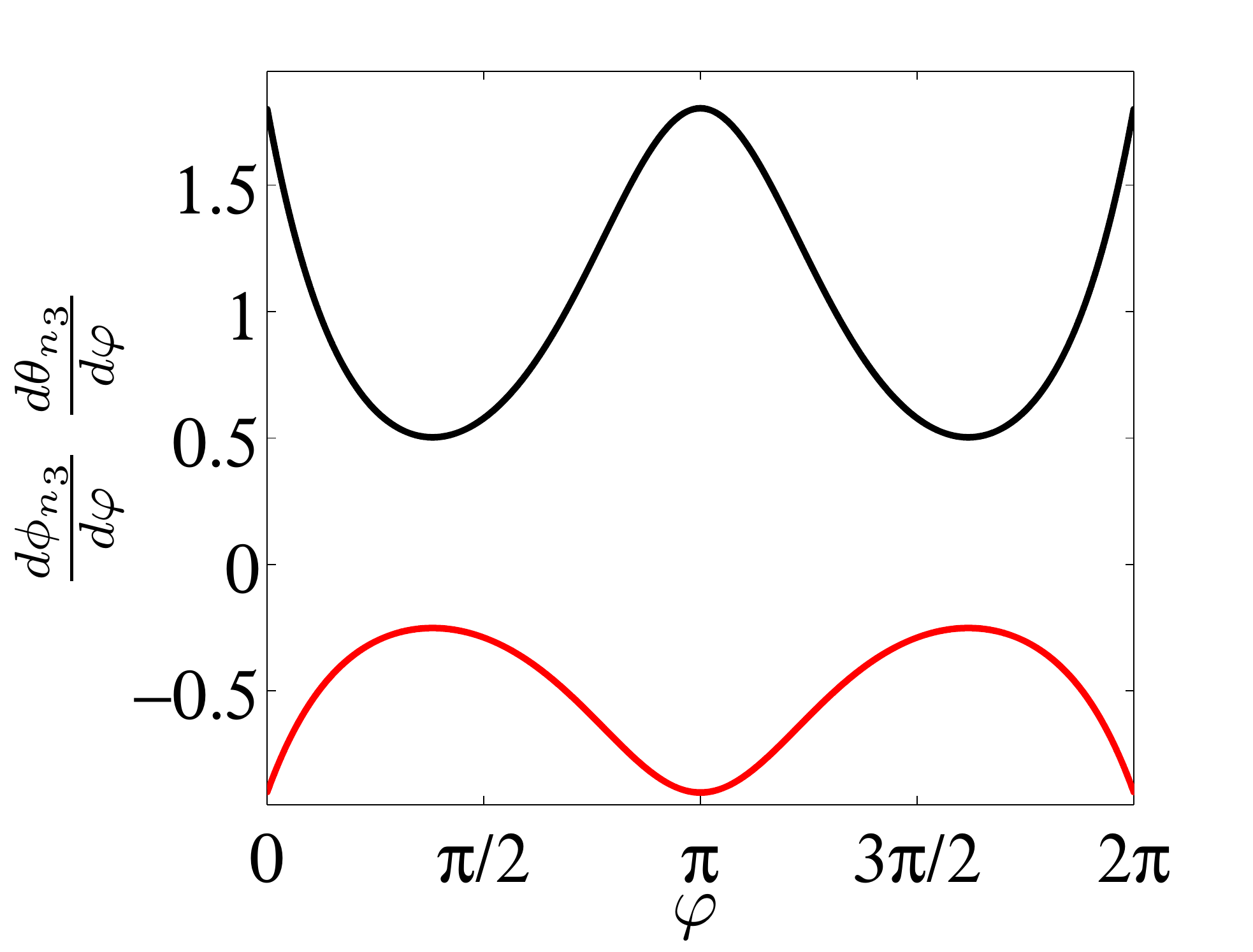}
\includegraphics[width=0.49\columnwidth]{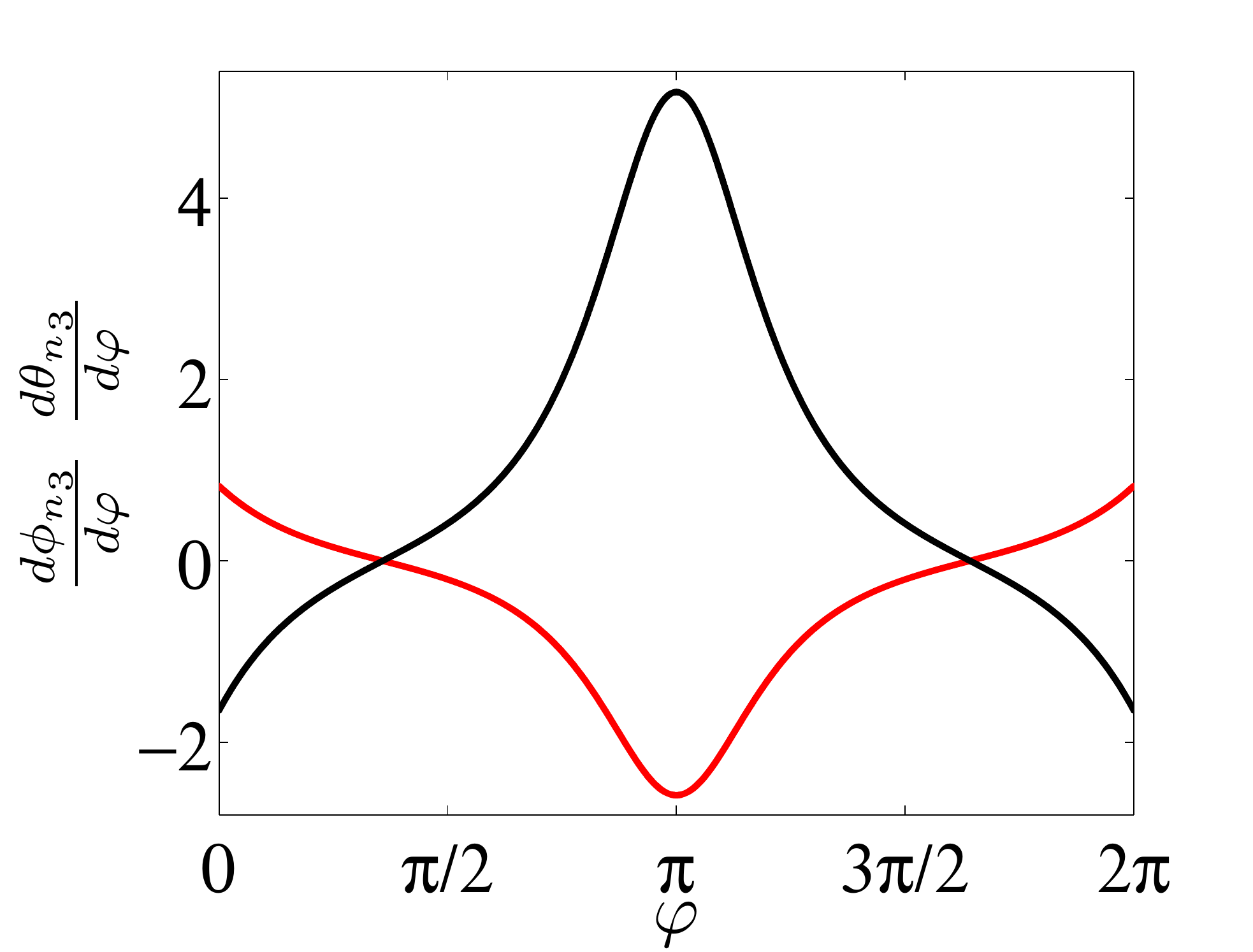}
\includegraphics[width=0.49\columnwidth]{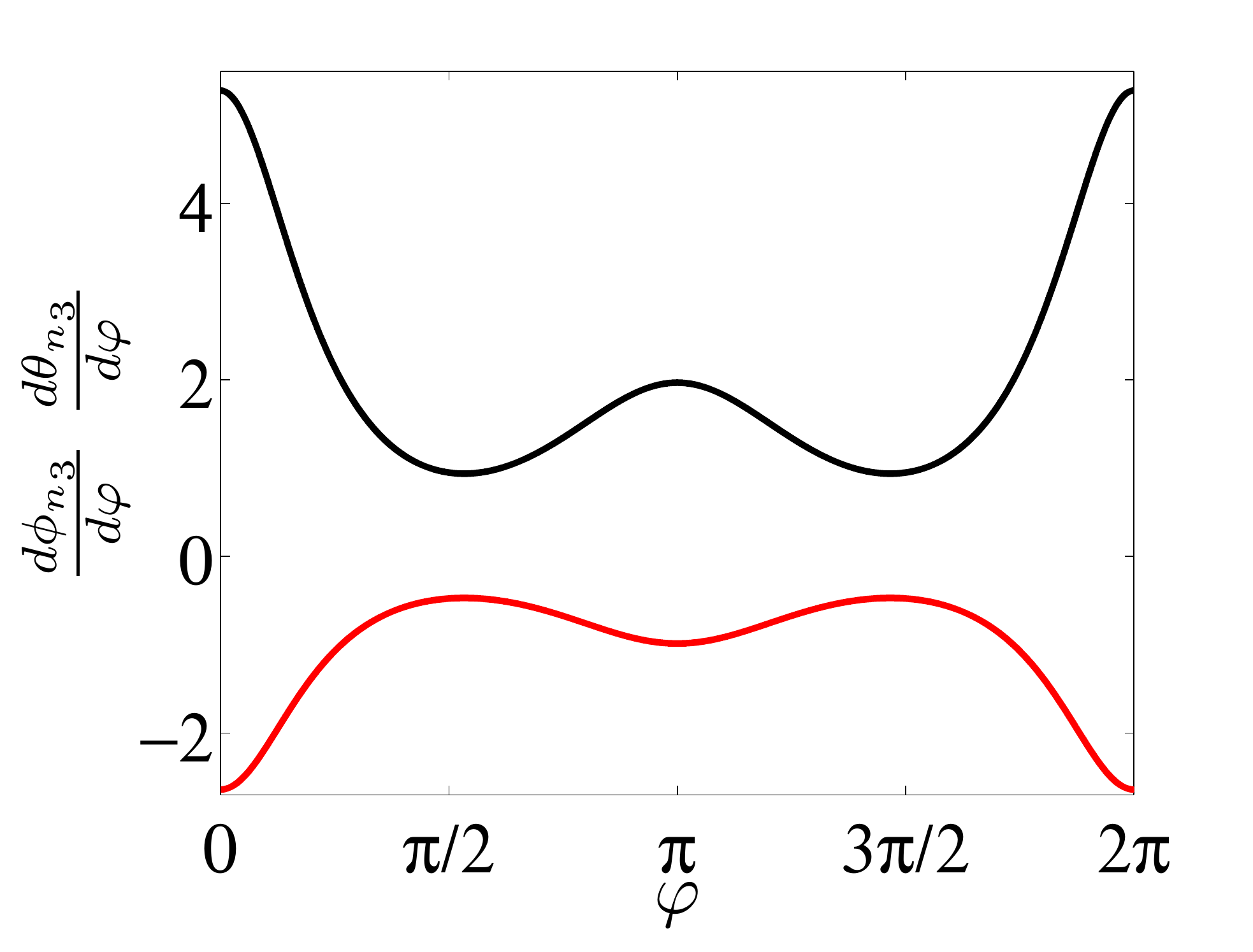}
\includegraphics[width=0.49\columnwidth]{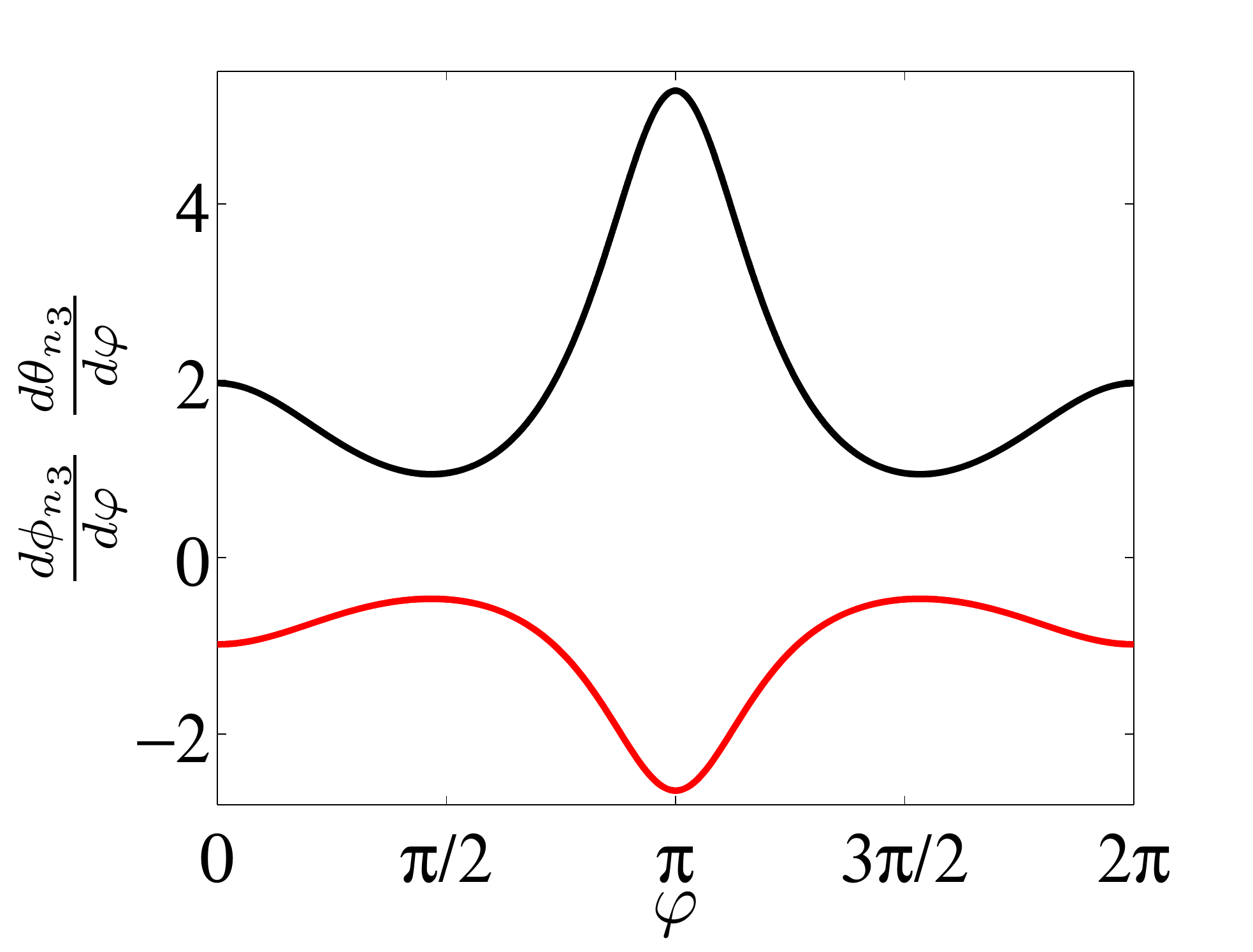}
\caption {Particular examples of two types of phase solitons for  $\phi$ (red line) and $\theta$  (black line) and for the winding number ${n_3} = 1$  (top) and ${n_3} = 2$  (bottom) in a BTRS three-band superconductor with one repulsive interband interaction. Here ${k_3} = 0.5$  has been adopted.}
\label{Examples_solitons}
\end{figure}

Generally speaking. there are two types of solutions of the Eqs. (\ref{FI_1}) and (\ref{FI_2}) dependent on the value of the constant $C \equiv {C_{{n_3}}}$, namely for $C \in \left[ { - 3{K_3},\frac{3}{2}{K_3}} \right]$ , $C \in \left( {\frac{3}{2}{K_3},\infty } \right)$. For the sake of simplicity, we provide the solutions for the interval values of the constant $C \in \left( {\frac{3}{2}{K_3},\infty } \right)$  respectively. Other type of solutions can be found by means of Jacobi imaginary transformations and the generalization of Landen’s transformations for the complex modulus \cite{Khare, Walker} of Jacobi elliptic functions.

\section{Particular examples of phase solitons}

In Figure \ref{Examples_solitons}, we present several different phase solitons for a BTRS three-band superconductor with one repulsive interband interaction, based on Eqs. (\ref{theta_sol_new1}) and (\ref{phi_sol_new1}) in the main paper. For larger clarity, we plot the derivatives of ${\theta _{{n_3}}}$  and ${\phi _{{n_3}}}$ .

\end{widetext}


\begin{thebibliography}{99}
\bibitem{remark}
Strictly speaking, at variance with the other examples mentioned below, these two pnictide materials should be considered with some caution because there at least one interband coupling is not weak. Nevertheless we hope that our results might be qualitatively correct in such a case, too.
\end{thebibliography}
\end{document}